\documentclass{revtex4}
\usepackage{amsmath}
\usepackage{amssymb}
\usepackage{amsfonts}
\usepackage{bm}
\usepackage[dvips]{graphicx}

\begin{document}

\title{Relaxation Functions of Ornstein-Uhlenbeck Process with Fluctuating Diffusivity}

\author{Takashi Uneyama}
\affiliation{%
  Center for Computational Science, Graduate School of Engineering,
  Nagoya University, Furo-cho, Chikusa, Nagoya 464-8603,
  Japan
}%

\author{Tomoshige Miyaguchi}
\affiliation{%
  Department of Mathematics, Naruto University of Education, Tokushima 772-8502,
  Japan}

\author{Takuma Akimoto}
\affiliation{%
Department of Physics, Tokyo University of Science, Noda, Chiba 278-8510, Japan
}%

\begin{abstract}
  We study a relaxation behavior of an Ornstein-Uhlenbeck {(OU)} process with a
  time-dependent and fluctuating diffusivity. In this process, the dynamics of a
  position vector is modeled by the Langevin equation with a linear restoring
  force and a fluctuating diffusivity { (FD)}.  This process can be interpreted as a
  simple model of the relaxational dynamics with internal degrees of freedom or in
  a heterogeneous environment.
 By utilizing the functional integral expression
  and the transfer matrix method, we show that the relaxation function can be
  expressed in terms of the eigenvalues and eigenfunctions of the transfer
  matrix, for general FD processes.
We apply our general theory to two simple FD
processes, where the FD is described by the Markovian
two-state model or an OU type process.
 We show
  analytic expressions of the relaxation functions in these models, and their
  asymptotic forms. We also show that the relaxation behavior of the
 OU process with an FD is qualitatively
  different from those obtained from conventional models such as the
  generalized Langevin equation.
\end{abstract}

\maketitle

%

\section{Introduction}
\label{introduction}

Recently, diffusion of Brownian particles with time-dependent and
fluctuating diffusivity has been studied extensively in many contexts such
as diffusion in heterogeneous environments and a simple model of normal yet
non-Gaussian
diffusion\cite{Chubynsky-Slater-2014,Bressloff-Newby-2014,Bressloff-2016,Massignan-Manzo-TorrenoPina-GarciaParajo-Lewenstein-Lapeyre-2014,Yamamoto-Akimoto-Hirano-Yasui-Yasuoka-2014,Manzo-TorrenoPina-Massignan-Lapeyre-Lewenstein-Parajo-2015,Uneyama-Miyaguchi-Akimoto-2015,Miyaguchi-Akimoto-Yamamoto-2016,Chechkin-Seno-Metzler-Sokolov-2017,Jain-Sebastian-2017,Serge-Bertaux-Rigneault-Marguet-2008}.
{ (The anomalous diffusion behavior of time-dependent
diffusivity models has been also studied\cite{Cherstvy-Metzler-2015,Cherstvy-Metzler-2016}.)}
The Langevin equation with a time-dependent and fluctuating diffusivity exhibits
non-trivial diffusion behavior.  If the system is in equilibrium, the
ensemble-averaged mean-square displacement (MSD) reduces to just a normal
diffusion which is the same as the simple Brownian motion.
However, this does not mean that the diffusion behavior
can be expressed as a normal and Gaussian process. If we consider the
{(time-averaged)} MSD data for individual realizations, the distribution of MSD data
around their ensemble-average does not obey the Gaussian distribution.
This is because the thermal
noise becomes non-Gaussian due to the multiplicative coupling between the
fluctuating diffusivity (FD) and the Gaussian noise.  
The non-Gaussian nature becomes evident if we consider the higher order correlation
functions.
For example, the authors showed that the relative standard deviation
of the time-averaged MSD (the square of which is also known as the ergodicity
breaking
parameter\cite{He-Burov-Metzler-Barkai-2008,Cherstvy-Chechkin-Metzler-2013}) can
be related to the correlation function of an FD\cite{Uneyama-Miyaguchi-Akimoto-2015}.

The concept of the FD seems to be useful to study the
microscopic dynamical behavior of molecules with the internal degrees of freedom
and/or in heterogeneous environments.
For example, protein dynamics has many internal degrees of freedom and the
conformational dynamics exhibits non-single-exponential type relaxation
behavior\cite{Yang-Luo-Karnchanaphanurach-Louie-Rech-Cova-Xun-Xie-2003,Yamamoto-Akimoto-Hirano-Yasui-Yasuoka-2014,Hu-Hong-Smith-Neusius-Cheng-Smith-2015}.
The diffusivity of the center of mass of an entangled polymer depends on the
end-to-end vector of the polymer
chain\cite{Doi-Edwards-1978,Doi-Edwards-book,Uneyama-Miyaguchi-Akimoto-2015}.
A molecule in a supercooled liquids exhibits heterogeneous and
fluctuating diffusion
behavior, which is known as the dynamic heterogeneity\cite{Yamamoto-Onuki-1998,Yamamoto-Onuki-1998a,Sillescu-1999}.  We
expect that other dynamical behavior, such as the relaxation behavior,
in these systems reflects their FDs.

Unfortunately, from the view point of experiments, the direct
observations of MSDs of individual molecules are not easy in some systems.
Instead of the direct observations of individual molecules, the
macroscopic measurements of relaxation functions (response
functions) are useful. The macroscopic relaxation functions can be
measured by imposing an external perturbation to the system and monitoring the
response of the system. The linear response theory gives the relation between
the time-correlation function of microscopic variables and the macroscopic
response function\cite{Evans-Morris-book}. Thus we can study the microscopic
{ molecular-level} dynamics from the macroscopic response functions.  For example,
we can investigate microscopic molecular dynamics of polymers from the
experimental data of the dielectric relaxation function (the response function
of the electric flux to the imposed electric field) and the relaxation modulus
(the response function of the stress to the imposed
strain)\cite{Watanabe-1999,Matsumiya-Uno-Watanabe-Inoue-Urakawa-2011}.

{ At the molecular-level, we expect that relaxation
dynamics can be described by a Langevin equation with a restoring force.
The Ornstein-Uhlenbeck (OU) process\cite{vanKampen-book} is the simplest
model to describe such dynamics. For example, in the harmonic
dumbbell model for a polymer, the dynamics is described as an OU type process.
Dynamics of a colloidal particle under an optical tweezer can also be
modeled as an OU type process.
If a molecule or a particle is in a heterogeneous environment
(such as a supercooled liquid), we expect that the relaxation behavior
is affected by an FD arising from the heterogeneity.
The relaxation behavior of a polymer or
a colloidal particle in a supercooled environment will be affected both by
a restoring force and a heterogeneous environment.
Dynamics of a protein may also be interpreted as relaxation dynamics
under a heterogeneous environment.}

{ Thus} we expect that the relaxation behavior of the systems with
FDs would be informative to model and/or
analyze microscopic molecular dynamics of polymers and supercooled liquids.
However, the relaxation behavior of the Langevin equation with an FD has not been studied in detail, as far as the authors know.  In this
work, we consider an OU type process with a
time-dependent and fluctuating diffusivity as a simple and analytically tractable
relaxation model with an FD.  We theoretically analyze
relaxation functions and show that the relaxation functions are strongly
affected by the dynamics of the diffusivity. We give a formal expression for the
relaxation function for a general Markovian dynamics of diffusivity. We show
that the form of the relaxation function is determined as a result of the
competition between the relaxation (diffusion) dynamics in the OU process
and the transition dynamics of the FD. Then we apply our result
to two simple and solvable models. One is the Markovian two-state model where the diffusivity can take only
two different values. Another is the OU type model where the
noise coefficient obeys an OU process.  We discuss the
properties of the relaxation functions by the OU process with an
FD, and possible applications of our theoretical results to
the analyses of experimental data.

\section{Model}
\label{model}

We consider the dynamics of a position in a $d$-dimensional space, $\bm{r}$. This
position can be interpreted as the position of a tagged particle in the system,
or a bond vector which connects two particles (in the dumbbell
model\cite{Kroger-2004}). In both cases, at the mesoscopic scale, the dynamics
of the position can be reasonably described by the Langevin type
equation.
We consider the dynamics of a particle or a bond vector in a
heterogeneous environment, where the diffusivity (or the mobility) fluctuates in
time. The effect of the inertia term in a dynamic
equation is generally
small at the mesoscopic scale, and thus we can safely ignore the inertial term (the overdamped limit).
As a simple model to describe such systems,
we employ the Langevin equation with an FD\cite{Uneyama-Miyaguchi-Akimoto-2015}:
\begin{equation}
 \label{langevin_equation_fluctuating_diffusivity}
  \frac{d\bm{r}(t)}{dt} = \frac{D(t) }{k_{B} T} \bm{F}(\bm{r},t)
  + \sqrt{2 D(t)} \bm{w}(t) .
\end{equation}
Here, $k_{B}$ is the Boltzmann constant, $T$ is the temperature, $D(t)$
is a time-dependent and fluctuating diffusivity, $\bm{F}(\bm{r},t)$ is the
force for the position, and $\bm{w}(t)$ is the Gaussian white noise.
We limited ourselves to the case where the diffusion coefficient is given as a
scalar quantity. (In general, the diffusion coefficient is a tensor quantity\cite{Uneyama-Miyaguchi-Akimoto-2015,Miyaguchi-2017}.)
From the fluctuation-dissipation relation of the second kind, the
first and second moments of the Gaussian noise $\bm{w}(t)$ are
{ $\langle \bm{w}(t) \rangle = 0$ and
$\langle \bm{w}(t) \bm{w}(t') \rangle = \bm{1} \delta(t - t')$,}
where $\langle \dots \rangle$ represents the statistical average and $\bm{1}$ is
the unit tensor in $d$-dimensions. We assume that the dynamics of $D(t)$ is
statistically independent of the noise $\bm{w}(t)$. Namely, the
dynamics of $D(t)$ is statistically not affected by the dynamics of
$\bm{r}(t)$.

\begin{figure}[h]
\includegraphics[width=.5\linewidth,clip]{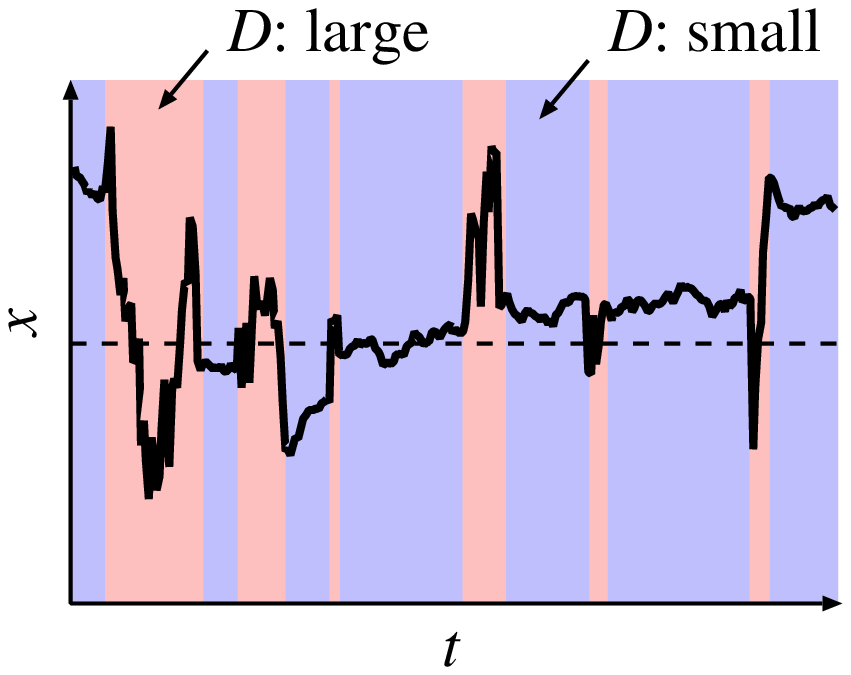}
\caption{\label{oufd_image}}
\end{figure}

Eq~\eqref{langevin_equation_fluctuating_diffusivity} can be interpreted
as a simplified model for dynamics of supercooled polymer
melts\cite{Baschnagel-Varnik-2005,Ding-Sokolov-2006} or trapped Brownian particles in
heterogeneous
environments\cite{Wong-Gardel-Reichman-Weeks-Valentine-Bausch-Weitz-2004,Shundo-Mizuguchi-Miyamoto-Goto-Tanaka-2011,Hori-Penaloza-Shundo-Tanaka-2012}.
For the case of a polymer chain in a supercooled melt, we consider the end-to-end vector as
the bond vector and the effective potential can be expressed as
a harmonic potential\cite{Doi-Edwards-book}.
For the case of a trapped Brownian particle,
we consider the situation where the Brownian particle is trapped by an optical
tweezer. The effective potential by an optical tweezer can be
approximated well by a harmonic potential\cite{Florin-Pralle-Stelzer-Horber-1998}.
Thus we assume that the force in
eq~\eqref{langevin_equation_fluctuating_diffusivity} is derived from a harmonic
potential
{ $U(\bm{r}) = \alpha k_{B} T \bm{r}^{2} / 2$, }
where $\alpha$ is constant. (If the force is purely entropic, $\alpha$ is
independent of the temperature $T$\cite{Doi-Edwards-book}. However, in general, $\alpha$ depends on the
temperature. Fortunately, the explicit form of $\alpha$ and its temperature
dependence is not important for the analyses in this work.)  The force
becomes
{ $\bm{F} = - {\partial U(\bm{r})} / {\partial \bm{r}} = - \alpha k_{B}
  T \bm{r}$ ,}
and eq~\eqref{langevin_equation_fluctuating_diffusivity} reduces to the
following OU process with an FD (OUFD):
\begin{equation}
 \label{ornstein_uhlenbeck_fluctuating_diffusivity}
 \frac{d \bm{r}(t)}{dt} = - \alpha D(t) \bm{r}(t)
  + \sqrt{2 D(t)} \bm{w}(t) .
\end{equation}
{We show an example of the OUFD process in Figure~\ref{oufd_image}.}
If $D(t)$ is constant, eq \eqref{ornstein_uhlenbeck_fluctuating_diffusivity}
reduces to a usual OU process\cite{vanKampen-book}, and we can
easily analyze it. {(As clearly observed in
Figure~\ref{oufd_image}, the fluctuation of the diffusivity
qualitatively affects the stochastic process, and thus generally the
OUFD behaves in a different way from a usual OU process.)}

To fully describe the OUFD, we need the dynamic equation for $D(t)$. We assume that
the dynamics of $D(t)$ obeys a Markovian stochastic process, and $D(t)$ is
sampled from the equilibrium ensemble. (The effect of non-equilibrium initial
ensembles to the OU process with a constant diffusivity was
recently studied\cite{Cherstvy-Thapa-Mardoukhi-Chechkin-Metzler-2018}.)  For
example, we can employ a Markovian jump processes between several discrete
states for the diffusivity, or a Langevin equation for the diffusivity. In any
case, as far as the process is Markovian, we can formally express the
time-evolution equation for the probability distribution of $D$ by a master
equation as
\begin{equation}
 \label{master_equation_diffusivity}
 \frac{\partial P(D,t)}{\partial t} = \hat{\mathcal{L}} P(D,t) ,
\end{equation}
where $P(D,t)$ is the probability distribution of $D$ at time $t$, and
$\hat{\mathcal{L}}$ is a linear operator (such as the Fokker-Planck operator and
the transition matrix\cite{vanKampen-book}). Or, more generally, we
  assume that the diffusivity $D(t)$ is obtained from a Markovian variable
  $\xi(t)$ as $D(t) = D(\xi(t))$, and that its probability distribution follows
  a master equation
\begin{equation}
 \label{master_equation_xi}
 \frac{\partial P(\xi,t)}{\partial t} = \hat{\mathcal{L}} P(\xi,t).
\end{equation}
Note that Eq.~\eqref{master_equation_diffusivity} is a special case for which
$D(t) = \xi(t)$.

To study the relaxational behavior of the OU process, we
consider the following relaxation function:
\begin{equation}
 \label{relaxation_function_phi_definition}
 \Phi(t) \equiv \frac{\langle \bm{r}(t) \cdot \bm{r}(0) \rangle}{\langle
  \bm{r}^{2}(0) \rangle} .
\end{equation}
Here, it should be noticed that the statistical average $\langle \dots \rangle$
is taken both for the noise $\bm{w}(t)$ and the diffusion coefficient
$D(t)$.
{
In this work, we take the statistical average of $\bm{w}(t)$ before
that of $D(t)$. The order of two statistical avareges does not affect
the results in most cases, including the equilibrium systems.}
Eq~\eqref{relaxation_function_phi_definition} can be interpreted as the
normalized relaxation function of $\bm{r}(t)$. If we interpret $\bm{r}(t)$ as
the bond vector and it has an electric dipole,
eq~\eqref{relaxation_function_phi_definition} can be related to the dielectric
relaxation function\cite{Watanabe-1999}. If
we interpret $\bm{r}(t)$ as the position of a trapped particle,
eq~\eqref{relaxation_function_phi_definition} can be related to the
ensemble-averaged MSD as
\begin{equation}
 \langle [\bm{r}(t) - \bm{r}(0)]^{2} \rangle
  = 2 \langle \bm{r}^{2}(0) \rangle [ 1 - \Phi(t) ].
\end{equation}
Therefore, if the relaxation function $\Phi(t)$ exhibits some non-trivial
properties, we will observe them via the ensemble-averaged MSD. This is in
contrast to the case without a trap potential, where the ensemble-averaged MSD
exhibits just a normal diffusion.

We calculate the relaxation function $\Phi(t)$ for the OUFD, from
eq~\eqref{ornstein_uhlenbeck_fluctuating_diffusivity} and \eqref{relaxation_function_phi_definition}. By integrating
eq~\eqref{ornstein_uhlenbeck_fluctuating_diffusivity} from time $0$ to $t$, we
have
\begin{equation}
 \label{ornstein_uhlenbeck_fluctuating_diffusivity_formal_solution}
  \begin{split}
 \bm{r}(t) & = \exp\left[ - \alpha \int_{0}^{t} dt' \, D(t')
		\right] \bm{r}(0) \\
   & \qquad + \int_{0}^{t} dt' \exp\left[ - \alpha \int_{t'}^{t} dt'' \, D(t'')
  \right] \sqrt{2 D(t')} \bm{w}(t') .
\end{split}
\end{equation}
Because $D(t)$, $\bm{w}(t)$ (for $t > 0$) and $\bm{r}(0)$ are statistically
independent of each other, we { take the average over
$\bm{w}(t)$ and $\bm{r}(0)$. Then we} have
\begin{equation}
 \langle \bm{r}(t) \cdot \bm{r}(0) \rangle
  = \left\langle \exp\left[ - \alpha \int_{0}^{t} dt' \, D(t')
		\right] \right\rangle \langle \bm{r}^{2}(0) \rangle ,
\end{equation}
and
\begin{equation}
 \label{relaxation_function_phi}
 \Phi(t) = \left\langle \exp\left[ - \alpha \int_{0}^{t} dt' \, D(t')
		\right] \right\rangle .
\end{equation}

For comparison, we also consider another relaxation function defined as
\begin{equation}
 \label{relaxation_function_psi_definition}
 \Psi(t) \equiv \frac{\langle [r_{x}(t) r_{y}(t)][r_{x}(0) r_{y}(0)]
  \rangle}{\langle [r_{x}(0) r_{y}(0)]^{2} \rangle} .
\end{equation}
We have assumed that the dimension of the space is at least $2$ ($d \ge 2$).
Eq~\eqref{relaxation_function_psi_definition} can be interpreted as the normalized relaxation
function of the off-diagonal component of a second rank tensor
$\bm{r}(t)\bm{r}(t)$. If we interpret $\bm{r}(t)$ as a bond vector, the
relaxation function $\Psi(t)$ can be related to the relaxation modulus
(the viscoelastic relaxation function)\cite{Watanabe-1999}.
The relaxation function $\Psi(t)$ can be calculated in a similar way to
$\Phi(t)$. From eq
\eqref{ornstein_uhlenbeck_fluctuating_diffusivity_formal_solution},
we have
\begin{equation}
 \langle r_{x}(t) r_{y}(t) r_{x}(0) r_{y}({0}) \rangle
  = \left\langle \exp\left[ - 2 \alpha \int_{0}^{t} dt' \, D(t')
		\right] \right\rangle \langle r_{x}^{2}(0)r_{y}^{2}(0) \rangle ,
\end{equation}
and thus
\begin{equation}
 \label{relaxation_function_psi}
 \Psi(t) = \left\langle \exp\left[ - 2 \alpha \int_{0}^{t} dt' \, D(t')
		\right] \right\rangle .
\end{equation}
If $D(t)$ is constant, $\Psi(t)$ can be related to $\Phi(t)$ simply as
$\Psi(t) = \Phi(2 t) = \Phi^{2}(t) $. Thus the two relaxation functions
are essentially the same. However, in the case of the OUFD, the relation between $\Psi(t)$
and $\Phi(t)$ is generally not that simple.

So far, the calculation of the relaxation functions $\Phi(t)$ and $\Psi(t)$ was rather
straightforward [eqs~\eqref{relaxation_function_phi} and \eqref{relaxation_function_psi}].  However, to obtain the explicit expression of these
relaxation functions, we need to evaluate the ensemble averages of the
state-dependent relaxation functions, which seem not to be trivial.
Because the
relaxation functions $\Phi(t)$ and $\Psi(t)$ have almost the same form, in the
followings we mainly consider the relaxation function $\Phi(t)$ unless
explicitly stated. Once we have the analytic expression for $\Phi(t)$, one for
$\Psi(t)$ can be easily obtained by replacing $\alpha$ by $2 \alpha$.
From eq~\eqref{relaxation_function_phi}, the relaxation function can be
calculated if the statistics of the integral $\int_{0}^{t} dt' \, D(t')$ is
known. There are several different methods to evaluate
eq~\eqref{relaxation_function_phi}. In this work we will utilize the transfer
matrix
method\cite{Scalapino-Sears-Ferrell-1972,Krumhansl-Schrieffer-1975,Bishop-Krumhansl-1975},
but other methods can be employed as well.  For example, from the view point of
the renewal theory, eq~\eqref{relaxation_function_phi} can be related to the
statistics of the occupation time or the
magnetization\cite{Godreche-Luck-2001,Akimoto-Yamamoto-2016}. Thus we can
utilize the methods developed in the field of the renewal theory to analyze the
relaxation function.  The analyses based on the renewal theory will be published
elsewhere\cite{Miyaguchi-inpreparation}.
Also, the propagator for a free Brownian
particle with an FD has a similar form to
eq~\eqref{relaxation_function_phi}\cite{Jain-Sebastian-2017}.
Thus the analyses for a free Brownian motion would 
be also utilized to the analyses of the relaxation function, and vise versa.

\section{Theory}
\label{theory}

To evaluate the statistical average over an FD, we introduce
the path probability\cite{Kleinert-book} which gives the statistical weight for
a certain realization of $D(t)$ or $\xi(t)$. We express the diffusion coefficient as $D(t) =
D(\xi(t))$, where $\xi(t)$ is a Markovian stochastic process
[eq~\eqref{master_equation_xi}].  (The stochastic
process $\xi(t)$ can be both a continuum stochastic process and a discrete jump
process.)  The path probability is expressed as a functional of $\xi(t)$, and we
describe it as $\mathcal{P}[\xi]$. The relaxation function $\Phi(t)$
\eqref{relaxation_function_phi} can be
rewritten as
\begin{equation}
 \label{relaxation_function_phi_functional_integral}
 \Phi(t) = \int \mathcal{D}\xi \, \exp\left[ - \alpha \int_{0}^{t} dt' \, D(\xi(t'))
		\right] \mathcal{P}[\xi],
\end{equation}
where $\int \mathcal{D}\xi$ represents the functional integral (or the path
integral) over the stochastic variable $\xi(t)$
\cite{Kleinert-book,Swanson-book}. We assume that the measure of the functional
integral is determined appropriately so that the total probability becomes
unity.  Eq~\eqref{relaxation_function_phi_functional_integral} has the similar
form as the partition function for the Ginzburg-Landau {(GL)} model in a one
dimensional space\cite{Onuki-book}.
  { [In the GL model,
the free energy of a system is expressed as a functional of the order
parameter field $\psi(x)$, as $\mathcal{F}[\psi]$. Under a constant external field which is conjugate
to the order parameter, $h$, an extra term $\int dx \, h \psi(x)$ is
added to the free energy functional. The partition function under the
external field is expressed as $\mathcal{Z} = \int \mathcal{D}\psi \,
\exp[ - (h / k_{B} T) \int dx \, \psi(x) - \mathcal{F}[\psi] / k_{B}
T]$, which has the same form as eq~\eqref{relaxation_function_phi_functional_integral}.]}
In analogy to the GL model,
$\alpha$ and $D(\xi(t))$ can be interpreted as an external field and the order
parameter which is conjugate to the applied external field, respectively. (This
situation would be similar to the Martin-Siggia-Rose
formalism\cite{Martin-Siggia-Rose-1973}.)
This analogy leads us to employ
techniques developed for the GL model, such as the transfer matrix
method.

We rewrite the path probability as
{ $\mathcal{P}[\xi] = \exp\left[ - \mathcal{S}[\xi] \right]$, }
where $\mathcal{S}[\xi]$ is the dimensionless action
functional\cite{Seifert-2012}. (In analogy to the GL model, this
action functional can be interpreted as the free energy functional
without an applied external field.)  For
convenience, we consider the discretized process for the diffusivity. Namely, we
discretize time $t$ as $t_{j} = j \Delta$ (with $\Delta$ being the time step
size), and approximate the function $\xi(t)$ by the set of discrete
points $\xi_{j} \equiv \xi(t_{j})$. Due to the Markovian nature, the action
functional can be rewritten as
\begin{equation}
\mathcal{S}[\xi] \approx \Delta \sum_{j} \, s(\xi_{j + 1}, \xi_{j}),
\end{equation}
where $s(\xi_{j + 1},\xi_{j})$ is a function of $\xi_{j + 1}$ and $\xi_{j}$.
The functional integral can be also rewritten as
{ $\int \mathcal{D}\xi \dotsb \approx \int \prod_{j} d\xi_{j} \dotsb$ .}
For simplicity, we assume that $t / \Delta$ is a positive integer. Then we can
rewrite eq \eqref{relaxation_function_phi_functional_integral} as follows:
\begin{equation}
 \label{relaxation_function_phi_markovian}
  \begin{split}
    \Phi(t) & \approx \int \prod_{j=0}^{t / \Delta} d\xi_{j} \,
    \exp\left[
    - \Delta \sum_{j = 0}^{t / \Delta-1} \alpha D(\xi_{j})
    - \Delta \sum_{j=0}^{t / \Delta-1} \, s(\xi_{j + 1}, \xi_{j})
    \right] P(\xi_{0}). \\
    & = \int \prod_{j = 0}^{t / \Delta} d\xi_{j} \, \exp\left[ - \Delta
    \sum_{j = 0}^{t / \Delta-1} [s(\xi_{j + 1}, \xi_{j}) + \alpha D(\xi_{j})]
    \right] P(\xi_{0}).
  \end{split}
\end{equation}
Here $P(\xi_{0})$ is the probability distribution of $\xi$ at the initial state,
and is given as the equilibrium probability distribution: $P(\xi_{0}) =
P_{\text{eq}}(\xi_{0})$.
 To derive eq~\eqref{relaxation_function_phi_markovian}, we
  have performed the functional integration
  over $\xi(t')$ for $t' < 0$ and $t' > t$, since the relaxation
  function in eq~\eqref{relaxation_function_phi_functional_integral}
  depends on $\xi(t')$ only in the time range $0 \le t' \le t$. The functional integral over
  $\xi(t')$ for $t' > t$ becomes unity, and the functional integral over
  $\xi(t')$ for $t' < 0$ gives the initial probability distribution $P(\xi_{0})$.
The function $s(\xi,\xi')$ can be related
to the linear operator $\hat{\mathcal{L}}$ in eq
\eqref{master_equation_xi}. The formal solution of eq
\eqref{master_equation_xi} for the time interval $\Delta = t_{j +
  1} - t_{j}$ is
\begin{equation}
 \label{master_equation_diffusivity_formal_solution}
 P(\xi,t_{j + 1}) = e^{\Delta \hat{\mathcal{L}}} P(\xi,t_{j}) .
\end{equation}
On the other hand, the factor $e^{-\Delta s(\xi_{j + 1},\xi_{j})}$ represents
the transition probability from the state $\xi_{j}$ to the state $\xi_{j + 1}$
during the time interval $\Delta$:
\begin{equation}
 \label{transition_diffusivity_by_action}
  P(\xi_{j + 1},t_{j + 1}) = \int d\xi_{j} \, e^{-\Delta s(\xi_{j +
  1},\xi_{j})} P(\xi_{j},t_{j}) .
\end{equation}
By comparing eqs
\eqref{master_equation_diffusivity_formal_solution} and
\eqref{transition_diffusivity_by_action}, we have the following simple
relation between $s(\xi,\xi')$ and $\hat{\mathcal{L}}$:
\begin{equation}
 \label{relation_action_time_evolution_operator}
  \int d\xi' e^{- \Delta s(\xi,\xi')} P(\xi') = e^{\Delta \hat{\mathcal{L}}} P(\xi),
\end{equation}
where $P(\xi)$ is an arbitrary function.

Eq~\eqref{relation_action_time_evolution_operator} means that the action
functional can be calculated from the linear operator $\hat{\mathcal{L}}$.
Then, eq~\eqref{relaxation_function_phi_markovian} can be solved by utilizing the
transfer matrix
technique\cite{Scalapino-Sears-Ferrell-1972,Krumhansl-Schrieffer-1975,Bishop-Krumhansl-1975},
in a similar way.  It would be worth mentioning that a similar method is
employed by Bressloff and Newby\cite{Bressloff-Newby-2014,Bressloff-2016}, to
analyze the diffusion properties of a model with a time-dependent and
fluctuating diffusivity.  We introduce the transfer operator
$\hat{\mathcal{W}}$:
\begin{equation}
 \label{relation_transfer_operator_time_evolution_operator}
 \begin{split}
  e^{-\Delta \hat{\mathcal{W}}} P(\xi)
  & = \int d\xi' \,
 \exp\left[ - \Delta [s(\xi, \xi') + \alpha D(\xi') ]
     \right] P(\xi') \\
  & = e^{\Delta \hat{\mathcal{L}}} [e^{-\Delta \alpha D(\xi)} P(\xi)] ,
 \end{split}
\end{equation}
where we have utilized eq~\eqref{relation_action_time_evolution_operator}.
Since $\Delta$ is small, the exponential functions can be expanded into the
power series of $\Delta$. By keeping only the leading order terms, we have
\begin{equation}
 \label{transfer_operator_explicit}
 - \hat{\mathcal{W}} P(\xi)
 \approx  [\hat{\mathcal{L}} - \alpha D(\xi)] P(\xi).
\end{equation}
Therefore, we find that the transfer operator $\hat{\mathcal{W}}$ consists of
two contributions. One is the time-evolution operator for the diffusivity
$\hat{\mathcal{L}}$, and another is the diffusivity dependent decay factor $-
\alpha D$. We may call the former as the ``transition dynamics'' and the latter as
the ``relaxation dynamics''.  At the limit of $\Delta \to 0$,
eq~\eqref{transfer_operator_explicit} becomes exact. Then, from
eqs~\eqref{relaxation_function_phi_markovian} and
\eqref{relation_transfer_operator_time_evolution_operator}, we have the
following simple expression for the relaxation function $\Phi(t)$:
\begin{equation}
 \label{relaxation_function_markovian_phi_transfer_operator}
  \Phi(t)
    = \int d\xi \, e^{-t \hat{\mathcal{W}}} P_{\text{eq}}(\xi) .
\end{equation}
Eq~\eqref{relaxation_function_markovian_phi_transfer_operator} means that the
relaxation function $\Phi(t)$ is determined by the transfer operator
$\mathcal{W}$ and the equilibrium probability distribution of $\xi$.

To proceed the calculation, we introduce the $n$-th eigenvalue and eigenfunction of
$\hat{\mathcal{W}}$, $\lambda_{n}$ and $\psi_{n}(\xi)$:
\begin{align}
 \label{eigenequation_transfer_operator}
 \hat{\mathcal{\mathcal{W}}} \psi_{n}(\xi) = \lambda_{n} \psi_{n}(\xi) .
\end{align}
For simplicity, here we assume that the eigenvalues are ordered ascendingly
($\lambda_{n} \le \lambda_{m}$ if $n < m$).  We construct the basis set by the
eigenfunctions as
\begin{equation}
 \int d\xi \, \psi^{\dagger}_{n}(\xi) \psi_{m}(\xi) = \delta_{n m},
\end{equation}
where $\psi^{\dagger}_{n}(\xi)$ is the eigenfunction of the adjoint
operator of $\hat{\mathcal{W}}$. (In general, the transfer operator is
not self-adjoint and thus $\psi^{\dagger}_{n}(\xi)$ does not coincide
to $\psi_{n}(\xi)$. The eigenfunctions $\psi_{n}(\xi)$ and
$\psi^{\dagger}_{n}(\xi)$ form a biorthogonal basis set\cite{Risken-book}.)
 
Then eq~\eqref{relaxation_function_phi_markovian} can be rewritten in terms of
the eigenvalues and eigenfunctions.  We can rewrite eq
\eqref{relaxation_function_markovian_phi_transfer_operator} with the eigenvalues
and eigenfunctions as:
\begin{equation}
 \label{relaxation_function_markovian_phi_eigenmodes}
  \begin{split}
  \Phi(t)
    & = \int d\xi \, \sum_{n} e^{-t \lambda_{n}} \psi_{n}(\xi)
   \int d\xi' \, \psi_{n}^{\dagger}(\xi') P_{\text{eq}}(\xi') \\
    & = \sum_{n} \phi_{n} e^{-t \lambda_{n}} ,
  \end{split}
\end{equation}
where
\begin{equation}
 \label{relaxation_function_markovian_phi_intensities}
 \phi_{n} \equiv  \int d\xi \, \psi_{n}(\xi)
   \int d\xi' \, \psi_{n}^{\dagger}(\xi') P_{\text{eq}}(\xi') .
\end{equation}
From eq~\eqref{relaxation_function_markovian_phi_eigenmodes}, we conclude that
the relaxation function of the OUFD is given as the sum of single-exponential relaxations, in the case
where the dynamics of the diffusivity is described by a Markovian stochastic
process.  The relaxation rate and intensity of the $n$-th mode are $\lambda_{n}$
and $\phi_{n}$, respectively. The relaxation time of the $n$-th mode is simply
given as $\tau_{n} \equiv 1 / \lambda_{n}$.  At the long time region, only the
eigenmode with the smallest eigenvalue is dominant, and the relaxation function asymptotically
approaches to a single-exponential type relaxation. The longest relaxation time
is $\tau_{n}$ with the smallest $n$.  Intuitively, the relaxation rate
$\lambda_{n}$ is determined by the competition between the relaxation dynamics
of the OU process which is characterized by the operator $-
\alpha D(\xi)$ and the transition dynamics of the diffusivity which is
characterized by the operator $\hat{\mathcal{L}}$.

The relaxation function $\Psi(t)$ can be calculated in almost the same way. As
we mentioned, the expression for $\Psi(t)$ is obtained by replacing $\alpha$ in
one for $\Phi(t)$ by $2 \alpha$ [eqs~\eqref{relaxation_function_phi} and
\eqref{relaxation_function_psi}]. This can be done by employing the following
transfer operator instead of eq~\eqref{transfer_operator_explicit}:
{ $ - \hat{\mathcal{W}}' P(\xi)  =  [\hat{\mathcal{L}} - 2 \alpha D(\xi)] P(\xi)$.}
Or, if we have the analytical expressions for $\lambda_{n}$, $\psi_{n}(\xi)$,
and $\psi_{n}^{\dagger}(\xi)$, we obtain the eigenvalue and eigenfunctions
for $\hat{\mathcal{W}}'$ by simply replacing $\alpha$ in them by $2
\alpha$.

The number of relaxation modes is finite, if the diffusivity is a discrete
variable and the number of states is finite. This corresponds to the case of the
Markovian $N$-state model, where the dynamics of the diffusion coefficient is
described by a stochastic jump process between states. In this case, the
stochastic variable $\xi(t)$ can take only $N$ values, and the probability
distribution function $P(\xi)$ reduces to the probability distributions for discrete
states $P_{n}$ (with $n = 1, 2, \dots, N$ being the index for
the discrete state).  In the Markovian $N$-state model, the equilibrium probability
distribution is expressed by an $N$-dimensional vector as $P_{\text{eq},n}$, and the transition
matrix is an $N \times N$ matrix, $L_{n,n'}$. Thus the transfer operator is
also an $N \times N$ matrix and the number of eigenvalues is $N$. Then there is
$N$ relaxation modes (some modes may be degenerated, and intensities of some
modes may be zero), and the relaxation function $\Phi(t)$ becomes
\begin{equation}
 \label{relaxation_function_markovian_finite}
  \Phi(t)
    = \sum_{n = 1}^{N} \phi_{n} e^{-t \lambda_{n}} .
\end{equation}
In the simplest, Markovian two-state model ($N = 2$), we have only two
relaxation modes. We show the detailed calculations for the Markovian two-state
model in the next section.

\section{Discussions}
\label{discussions}

\subsection{Two-State Model}
\label{two_state_model}

As a simple yet non-trivial example, we consider a simple model where the
diffusivity obeys the Markovian two-state
model\cite{Sillescu-1999,Uneyama-Miyaguchi-Akimoto-2015} (where the stochastic
variable $\xi(t)$ can take only two values).  As we mentioned, there are only
two relaxation modes in this case [$N = 2$ in
eq~\eqref{relaxation_function_markovian_finite}].  Here we analyze the behavior
of the two-state model in detail. We describe two states in the model as the
fast ($f$) and the slow ($s$) states, and describe the probability distributions
of the fast and slow states as $P_{f}(t)$ and $P_{s}(t)$.  We describe the
diffusivity at the fast and slow states as $D_{f}$ and $D_{s}$ ($D_{f} \ge
D_{s}$).  Eq~\eqref{master_equation_xi} now reduces to the following
simple equation:
\begin{equation}
 \label{master_equation_diffusivity_two_state_model}
 \frac{d}{dt}
  \begin{bmatrix}
   P_{f}(t) \\
   P_{s}(t)
  \end{bmatrix}
  =
  \begin{bmatrix}
   -k_{f} & k_{s} \\
   k_{f} & - k_{s}
  \end{bmatrix}
  \cdot
  \begin{bmatrix}
   P_{f}(t) \\
   P_{s}(t)
  \end{bmatrix} ,
\end{equation}
where $k_{f}$ and $k_{s}$ are the transition rate from the fast to slow
states, and from the slow to fast states, respectively. The equilibrium
distribution is simply given as
\begin{equation}
 \label{equilibrium_distribution_two_state_model}
 P_{\text{eq},f} = \frac{k_{s}}{k_{f} + k_{s}}, \qquad
 P_{\text{eq},s} = \frac{k_{f}}{k_{f} + k_{s}} .
\end{equation}

{
As we showed in Sec.~\ref{theory}, this model has only two relaxation modes.
The explicit form of the relaxation function can be analytically
calculated as
\begin{equation}
 \label{relaxation_function_phi_two_state_model}
\begin{split}
  \Phi(t)  = \phi_{-} e^{-t
 \lambda_{-}} + \phi_{+} e^{-t \lambda_{+}} ,
\end{split}
\end{equation}
where $\lambda_{\pm}$ and $\phi_{\pm}$ are the relaxation rates and
intensities of two relaxation modes. Their explicit forms are given as
follows, with the relaxation rate defined as $\mu_{h}\equiv \alpha D_{h}$ ($h = f, s$):
\begin{equation}
 \label{eigenvalues_two_state_model}
 \lambda_{\pm} \equiv
  \frac{1}{2}
  \left[ \mu_{f} + \mu_{s} + k_{f} + k_{s}
  \pm \sqrt{(\mu_{f} + \mu_{s} + k_{f} + k_{s})^{2}
  - 4 (k_{f} \mu_{s} + k_{s} \mu_{f} + \mu_{f} \mu_{s})}
  \right] ,
\end{equation}
\begin{equation}
 \label{intensity_two_state_model}
 \phi_{\pm} \equiv \frac{1}{\lambda_{+} - \lambda_{-}} \left[ \pm \frac{ 
    k_{f} \mu_{s} +  k_{s} \mu_{f} }{k_{s} +
 k_{f} } \mp
 \lambda_{\mp} \right] . 
\end{equation}
(See Appendix~\ref{relaxation_modes_for_two_state_model} for the derivation.)
}
From {eq~\eqref{intensity_two_state_model}}, we have $\phi_{+} + \phi_{-} = 1$. (This
is trivial since the relaxation function $\Phi(t)$ is normalized.)

The relaxation times of two modes are given as $\tau_{\pm} \equiv 1 /
\lambda_{\pm}$. From eq~\eqref{eigenvalues_two_state_model}, the explicit
expressions of $1 / \lambda_{\pm}$ contain both the relaxation and transition
rates. This means that the relaxation times do not coincide to the relaxation
times naively estimated as the inverse relaxation rates $1 / \mu_{f}$ and $1 /
\mu_{s}$.  Also, from {eq~\eqref{intensity_two_state_model}}, the intensities for two modes are also
affected by both the relaxation and transition rates.  Therefore, we conclude
that the behavior of the relaxation function $\Phi(t)$ is generally not simple,
even for the simple Markovian two-state model.

Although the behavior of the relaxation function $\Phi(t)$ given by
eq~\eqref{relaxation_function_phi_two_state_model} is generally not simple, the
relaxation function reduces to simple forms at some special cases.  Here we
consider two limiting cases. The first case is the case where the transition
between the fast and slow states is sufficiently fast. In this case we assume that
$k_{f}, k_{s} \gg \mu_{f}, \mu_{s}$.  The relaxation rates reduce to
{
 $\lambda_{-} \approx (k_{f} \mu_{s} + k_{s} \mu_{f}) / (k_{f} + k_{s})$ and
 $\lambda_{+} \approx k_{f} + k_{s}$,
}
and the intensities reduce to
{ $ \phi_{-} \approx 1$ and $\phi_{+}  \approx 0 $.}
Therefore, in this case the second mode disappears and the
single-exponential type relaxation behavior is recovered:
\begin{equation}
 \label{relaxation_function_phi_two_state_model_fast_transition_limit}
 \Phi(t) \approx \exp\left( - \frac{k_{f} \mu_{s} + k_{s} \mu_{f}}{k_{f}
		      + k_{s}} t \right) .
\end{equation}
Eq~\eqref{relaxation_function_phi_two_state_model_fast_transition_limit} means
that the relaxation time is given as the harmonic average of the relaxation
times of the fast and slow states.  Intuitively, this result can be understood
as follows; due to the fast transition between the fast and slow states, the
diffusivity $D(t)$ can be replaced by the equilibrium average $\langle D
\rangle$. Then, the effective relaxation rate is estimated to be
\begin{equation}
 \alpha \langle D \rangle = \alpha (D_{f} P_{\text{eq},f} + D_{s} P_{\text{eq},f})
 = \frac{k_{f} \mu_{s} + k_{s} \mu_{f}}{k_{f} + k_{s}},
\end{equation}
and this coincides to $\lambda_{-}$.

The second case is the case where the transition between fast and slow states is
sufficiently slow. We assume that $k_{f}, k_{s} \ll \mu_{f}, \mu_{s}$,
and then the relaxation rates become
{ $\lambda_{-} \approx \mu_{s}$ and $\lambda_{+} \approx
\mu_{f}$, }
and the relaxation intensities are
{ $\phi_{-} \approx {k_{f}} / (k_{f} + k_{s})$ and
$\phi_{+}  \approx {k_{s}} / (k_{f} + k_{s})$.}
Therefore, the relaxation function $\Phi(t)$ consists of two modes with
the relaxation times determined solely by the relaxation rates
$\mu_{f}$ and $\mu_{s}$:
\begin{equation}
 \label{relaxation_function_phi_two_state_model_slow_transition_limit}
  \Phi(t) \approx \frac{k_{f}}{k_{f} + k_{s}} e^{-\mu_{s} t}
  +  \frac{k_{s}}{k_{f} + k_{s}} e^{-\mu_{f} t} .
\end{equation}
In this case, we have two relaxation modes and thus non-single-exponential type
behavior is observed. The relaxation times coincide to the relaxation times of
the pure fast and slow states, and the intensities correspond to the equilibrium
fractions, $\phi_{-} \approx P_{\text{eq},s}$ and $\phi_{+} \approx
P_{\text{eq},f}$ [eq\eqref{equilibrium_distribution_two_state_model}].
Intuitively, this case corresponds to the mixture of two
statistically independent relaxation processes. Because the transition rates are small, the
system can fully relax before the transition occurs. Thus the relaxation times
are not affected by the transition dynamics, and the intensities are just given
as the equilibrium probabilities.

\begin{figure}[h]
\includegraphics[width=.8\linewidth,clip]{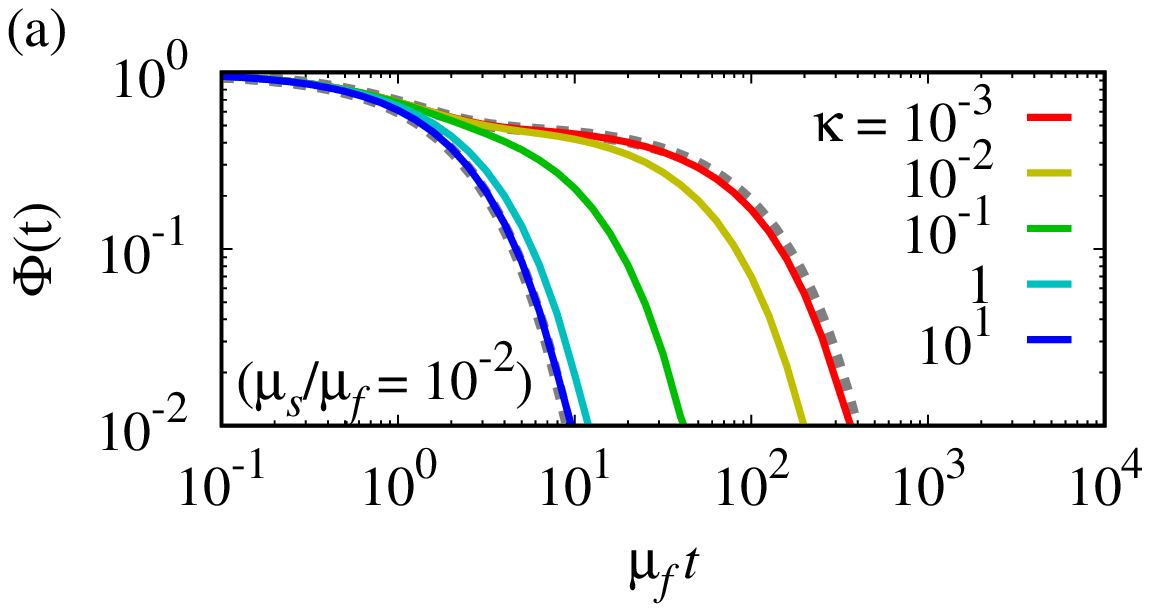}
\includegraphics[width=.8\linewidth,clip]{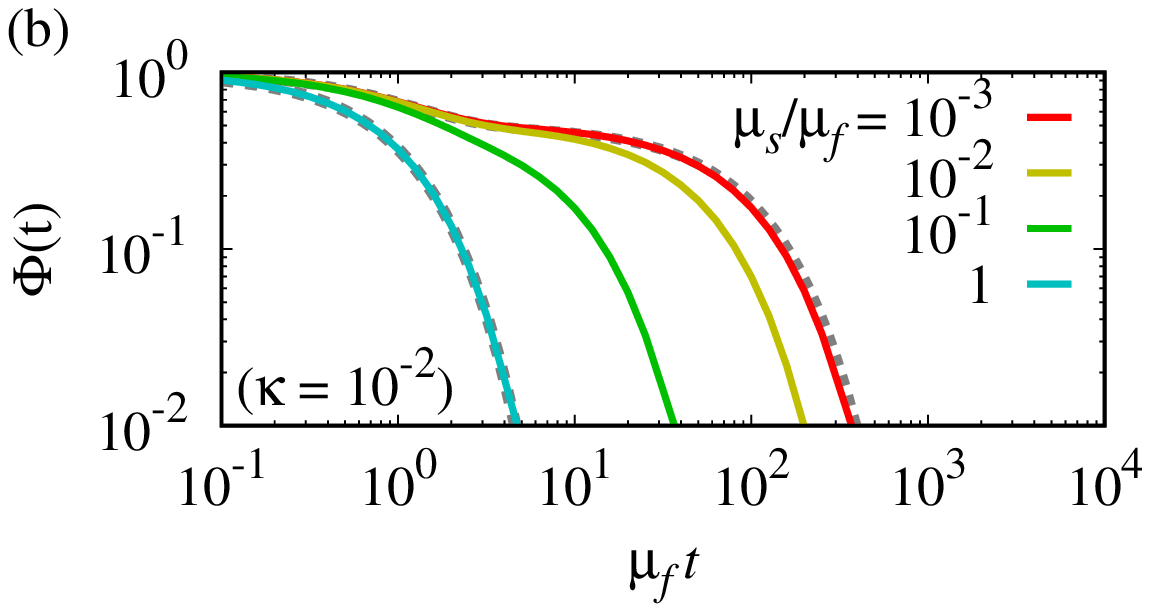}
\caption{\label{oufd_two_state_relaxation_function}}
\end{figure}

We show the relaxation function $\Phi(t)$ for various transition and relaxation
rates in Figure~\ref{oufd_two_state_relaxation_function}. For simplicity, here
we limit ourselves to the case where two transition rates are the same: $k_{f} =
k_{s}$. In this case, we have essentially two freely tunable parameters, $\kappa
\equiv k_{f} / \mu_{f} = k_{s} / \mu_{f}$ and $\mu_{s} / \mu_{f}$.
Figure~\ref{oufd_two_state_relaxation_function}(a) shows the relaxation function
for $\mu_{s} / \mu_{f} = 10^{-2}$ and various values of $\kappa$.  The
asymptotic forms for the fast and slow transition limits
[eqs~\eqref{relaxation_function_phi_two_state_model_fast_transition_limit} and
\eqref{relaxation_function_phi_two_state_model_slow_transition_limit}] are also
shown for comparison. We can observe that even if the value of $\mu_{s} /
\mu_{f}$ is constant, the relaxation function largely changes if we change
$\kappa$.  For small and large $\kappa$ cases, we observe that the asymptotic
forms work as good approximations.
Figure~\ref{oufd_two_state_relaxation_function}(b) shows the relaxation function
for $\kappa = 10^{-2}$ and various values of $\mu_{s} / \mu_{f}$. For the case
of $\mu_{s} / \mu_{f} = 1$, the relaxation function trivially reduces to a
single exponential form, $\Phi(t) = \exp(-\mu_{f} t)$. For the case of small
$\mu_{s}$, we consider the condition $\mu_{s} / \mu_{f} \ll \kappa \ll 1$ and
have $\lambda_{+} \approx \mu_{f}$, $\lambda_{-} \approx \mu_{f} \kappa$, and
$\phi_{+} \approx \phi_{-} \approx 1 / 2$.  We observe that the data for small
and large $\mu_{s} / \mu_{f}$ agree well with the asymptotic forms.

Except the special cases examined above, in general, the two relaxation times
cannot be simply related to the relaxation rates of the fast and slow states.
In addition, we cannot determine the relaxation and transition rates solely from
the relaxation function $\Phi(t)$, even if $\Phi(t)$ can be approximately
expressed as the sum of two relaxation modes.  To investigate whether the
relaxation behavior is really affected by the FD or not,
we can utilize another relaxation function $\Psi(t)$. The relaxation function
$\Psi(t)$ can be obtained by replacing $\alpha$ in $\Phi(t)$ by $2 \alpha$, as
we mentioned. Thus, for the current case, we have
\begin{equation}
 \label{relaxation_function_psi_two_state_model}
 \Psi(t) = \phi_{-}' e^{- t \lambda_{-}'}
  +  \phi_{+}' e^{- t \lambda_{+}'} ,
\end{equation}
with $\lambda_{\pm}'$ and $\phi_{\pm}'$ defined as
\begin{equation}
 \label{eigenvalues_two_state_model_prime}
 \lambda_{\pm}' \equiv
  \frac{1}{2}
  \left[ 2 \mu_{f} + 2 \mu_{s} + k_{f} + k_{s}
  \pm \sqrt{(2 \mu_{f} + 2 \mu_{s} + k_{f} + k_{s})^{2}
  - 8 (k_{f} \mu_{s} + k_{s} \mu_{f} + 2 \mu_{f} \mu_{s})}
  \right] ,
\end{equation}
{
\begin{equation}
 \label{intensity_two_state_model_prime}
 \phi_{\pm}'  \equiv \frac{1}{\lambda_{+}' - \lambda_{-}'} \left[ \pm \frac{ 
    2 (k_{f} \mu_{s} +  k_{s} \mu_{f}) }{k_{s} +
 k_{f} } \mp
 \lambda_{\mp}' \right] .
\end{equation}
}
We consider two limiting cases again.  For the case where $k_{f}, k_{s} \gg
\mu_{f}, \mu_{s}$, eq~\eqref{relaxation_function_psi_two_state_model} reduces to
a simple exponential form as:
\begin{equation}
 \label{relaxation_function_phi_psi_two_state_model_fast_transition_limit}
 \Psi(t) \approx e^{- 2 \alpha \langle D \rangle t} \approx \Phi(2 t) \approx \Phi^{2}(t) .
\end{equation}
Thus we find that in this case the relaxation behavior is the same as
the usual OU process with a constant diffusivity.
On the other hand, for $k_{f}, k_{s} \ll \mu_{f}, \mu_{s}$, the relaxation function
$\Psi(t)$ simply reduces
\begin{equation}
 \label{relaxation_function_phi_psi_two_state_model_slow_transition_limit}
 \Psi(t) \approx P_{\text{eq},s} e^{- 2 \mu_{s} t}
  + P_{\text{eq},s} e^{- 2 \mu_{f} t} \approx \Phi(2 t) \neq \Phi^{2}(t).
\end{equation}
This means that the relation between $\Phi(t)$ and $\Psi(t)$ becomes
different from one for the constant diffusivity case.
In general, the relation between $\Phi(t)$ and $\Psi(t)$ is not simple.
We conclude that by combining the two relaxation functions
$\Phi(t)$ and $\Psi(t)$, we are able to extract some information on the
FD such as the transition rates between states.
For example, if we have four relaxation times $1 / \lambda_{\pm}$
and $1 / \lambda_{\pm}'$ from relaxation functions, we can determine
relaxation and transition rates $\mu_{f}, \mu_{s}, k_{f},$ and $k_{s}$.

This analysis method will be useful to analyze relaxation functions obtained by
experiments. The dielectric relaxation functions obtained by the dielectric
measurements can be related to the relaxation function $\Phi(t)$. The relaxation
moduli obtained by rheological measurements can be related to the relaxation
function $\Psi(t)$.  Matsumiya et
al\cite{Matsumiya-Uno-Watanabe-Inoue-Urakawa-2011} reported both the rheological
and dielectric relaxation data for glassy polystyrene samples with various
molecular weights. They showed that the storage modulus and the dielectric
relaxation function of the same sample have almost the same form but the
relaxation times are different. Although their data cannot be simply expressed
as two relaxation modes, analyses based on our results would be informative to
understand the nature of glassy dynamics in polymers.

\subsection{Ornstein-Uhlenbeck Type Model}
\label{ornstein_uhlenbeck_type_model}

We consider another simple case, where the dynamics of the diffusivity obeys an
OU type process. While the diffusivity is a discrete variable in
the two-state model, the diffusivity is a continuum variable in this model. Thus
the relaxation function consists of infinite relaxation modes (at least
formally).

Since the diffusivity should be positive, we introduce the noise coefficient of
which square gives the diffusion coefficient:
\begin{equation}
 \label{diffusivity_and_noise_coefficient}
 D(t) = \bar{D} b^{2}(t),
\end{equation}
where $b(t)$ is the noise coefficient and can be both positive and negative, and $\bar{D}$ is constant.
We interpret $b(t)$ as the stochastic variable
$\xi(t)$ in eqs~\eqref{master_equation_xi},
\eqref{relaxation_function_markovian_phi_transfer_operator} and
\eqref{relaxation_function_markovian_phi_eigenmodes}.  For the dynamics of
$b(t)$, we employ the following Langevin equation:
\begin{equation}
 \label{ornstein_uhlenbeck_noise_coefficient}
 \frac{d b(t)}{dt} = - k b(t) + \sqrt{2 k} w'(t) .
\end{equation}
Here $k$ is the rate constant and $w'(t)$ is the Gaussian white noise of
which the first and second moments are given as
{ $\langle w'(t) \rangle = 0$ and
$\langle w'(t) w'(t') \rangle = \delta(t - t')$.}
Eq~\eqref{ornstein_uhlenbeck_noise_coefficient} is an OU
process. A similar model for the noise coefficient was employed to model
  the diffusion behavior in a heterogeneous medium (the diffusing diffusivity
model)
\cite{Chechkin-Seno-Metzler-Sokolov-2017,Jain-Sebastian-2017,Tyagi-Cherayil-2017}.
{In the diffusing diffusivity model, the diffusion coefficient is
expressed as the square of a vector variable which obeys an OU
process. Our model can be interpreted as one-dimensional version of the
diffusing diffusivity model.}

{ Because the stochastic process for the diffusion coefficient is
fully specified, now we can calculate the explicit form of the
relaxation function $\Phi(t)$. Unlike the case of the two-state model,
the OU type model has infinite relaxation modes. The explicit expression
for the relaxation function is}
\begin{equation}
 \label{relaxation_function_noise_coefficient}
 \Phi(t) = \sum_{n = 0}^{\infty} \phi_{n}
 \exp\left[ - k  [ (2 n + 1) \gamma - 1 /2 ] t \right] ,
\end{equation}
{ where we have defined $\gamma \equiv \sqrt{\mu / k + 1 / 4}$
with $\mu \equiv \alpha \bar{D}$, and
\begin{equation}
 \label{intensity_noise_coefficient_explicit}
  \phi_{n} \equiv
\begin{cases}
 \displaystyle
 \frac{\sqrt{2} n!}{2^{n} [(n / 2)!]^{2}} 
  \frac{\gamma^{1/2}(\gamma - 1/2)^{n}}{(\gamma +  1/2)^{n + 1}} 
 & (n : \text{even}) , \\
 0 & (n : \text{odd}) .
\end{cases} 
\end{equation}
}
(The detailed calculations are shown in
Appendix~\ref{relaxation_modes_for_ornstein_uhlengeck_type_model}.)
From eq~\eqref{intensity_noise_coefficient_explicit}, only the modes with even $n$
survive. We set $n = 2 m$ and rewrite
eq~\eqref{relaxation_function_noise_coefficient} as
\begin{equation}
 \label{relaxation_function_noise_explicit}
 \Phi(t) = \sum_{m = 0}^{\infty} 
 \frac{\sqrt{2} (2m)!}{2^{2m} (m!)^{2}} 
  \frac{\gamma^{1/2}(\gamma - 1/2)^{2m}}{(\gamma +  1/2)^{2m + 1}} 
  \exp\left[ - k  [ (4 m + 1) \gamma - 1 /2 ] t \right] .
\end{equation}

The longest relaxation time $\tau_{0}$ is the inverse of the relaxation
rate for $n = 0$ by eq \eqref{eigenvalue_noise_coefficient}:
\begin{equation}
 \label{longest_relaxation_time_explicit}
 \tau_{0} \equiv \frac{1}{\lambda_{0}} 
  =  \frac{1}{k (\sqrt{\mu / k + 1/ 4} - 1/2)} .
\end{equation}
If the transition dynamics is much faster or slower than the relaxation
 dynamics, we have simple approximate forms for the
longest relaxation time:
\begin{equation}
 \label{longest_relaxation_time_explicit_asymptotic}
 \tau_{0} \approx
  \begin{cases}
   1 / \mu & (k \gg \mu), \\
   1 / \sqrt{k \mu} & (k \ll \mu) .
  \end{cases}
\end{equation}
Eq~\eqref{longest_relaxation_time_explicit_asymptotic} means that the
relaxation behavior of this model is largely affected by the transition
rate if the transition dynamics is slow. 
We consider two limiting cases in detail, as the case of the Markovian two-state
model. First, we assume that the transition dynamics is much faster than
the relaxation dynamics and assume $k \gg \mu$. 
{ In this case, we have
$\gamma \approx 1 / 2 + \mu / k$, and}
the intensity becomes
\begin{equation}
 \phi_{n} \approx
 \begin{cases}
  1 & (n = 0) , \\
  0 & (\text{otherwise}) .
 \end{cases}
\end{equation}
{ Therefore} the relaxation function reduces to the single-exponential form:
\begin{equation}
 \label{relaxation_function_noise_coefficient_fast_transition_limit}
 \Phi(t) \approx \exp( - \mu t) .
\end{equation}
From eq~\eqref{relaxation_function_noise_coefficient_fast_transition_limit}, the
relaxation time $1 / \mu$ is independent of the transition rate
$k$.
This is the same as the case of the Markovian two-state model. As
before, this result can be intuitively understood by considering the
equilibrium average of the relaxation rate:
\begin{equation}
 \alpha \langle D \rangle = \alpha \int_{-\infty}^{\infty} db \, \bar{D} b^{2}
  P_{\text{eq}}(b) = \mu ,
\end{equation}
{ where $P_{\text{eq}}(b) = e^{-b^{2} / 2} / \sqrt{2 \pi}$ is the
equilibrium distribution for the noise coefficient.}

Second, we assume that the transition dynamics is much slower than the
relaxation dynamics. In this case we have $k \ll \mu$ and $\gamma \gg 1$, but it is rather
difficult to calculate the approximate form for the relaxation function
from eq~\eqref{relaxation_function_noise_explicit} under this condition.
{Fortunately, we can calculate the approximate form starting
from the dynamic equation. The result is
\begin{equation}
 \label{relaxation_function_noise_coefficient_slow_transition_limit}
 \Phi(t) \approx \frac{1}{\sqrt{1 + 2 \mu t}} .
\end{equation}
(See
Appendix~\ref{calculation_of_relaxation_function_at_slow_transitoin_limit}
for detailed calculations.)}

Thus, in this case, we observe the power-law type behavior at the long time
region, $\Phi(t) \propto t^{-1/2}$ ($t \gtrsim 1 / 2 \mu$). Such power-law type
relaxation behavior is also observed for polymers (the Rouse
model)\cite{Doi-Edwards-book} and critical gels \cite{Winter-Chambon-1986}.  In
most cases, the power-law type relaxation is interpreted as the relaxation of
fractal structures where the relaxation time distribution
is given as a power-law type distribution. Our result gives another interpretation; the power-law type
relaxation can also be attributed to the FD.  As the case
of the two-state model, intuitively, the relaxation function is expressed as the
sum of relaxation modes and their intensities are given as the equilibrium
probability distribution.  In the same way,
eq~\eqref{relaxation_function_noise_coefficient_slow_transition_limit} can be
reproduced as the average of the relaxation function with respect to the
equilibrium distribution of the noise coefficient:
\begin{equation}
 \label{relaxation_function_noise_coefficient_no_transition}
 \Phi(t) 
 \approx \int_{-\infty}^{\infty} db \, e^{- \alpha \bar{D} b^{2} t} P_{\text{eq}}(b) 
  = \frac{1}{\sqrt{1 + 2 \mu t}} .
\end{equation}
The same expression as
eq~\eqref{relaxation_function_noise_coefficient_no_transition} can be
obtained by substituting {the approximate transfer operator
[eq~\eqref{transfer_operator_noise_coefficient_slow_transition_limit}
in Appendix~\ref{calculation_of_relaxation_function_at_slow_transitoin_limit}]}
directly into eq~\eqref{relaxation_function_markovian_phi_transfer_operator}.

\begin{figure}[h]
 \includegraphics[width=.8\linewidth,clip]{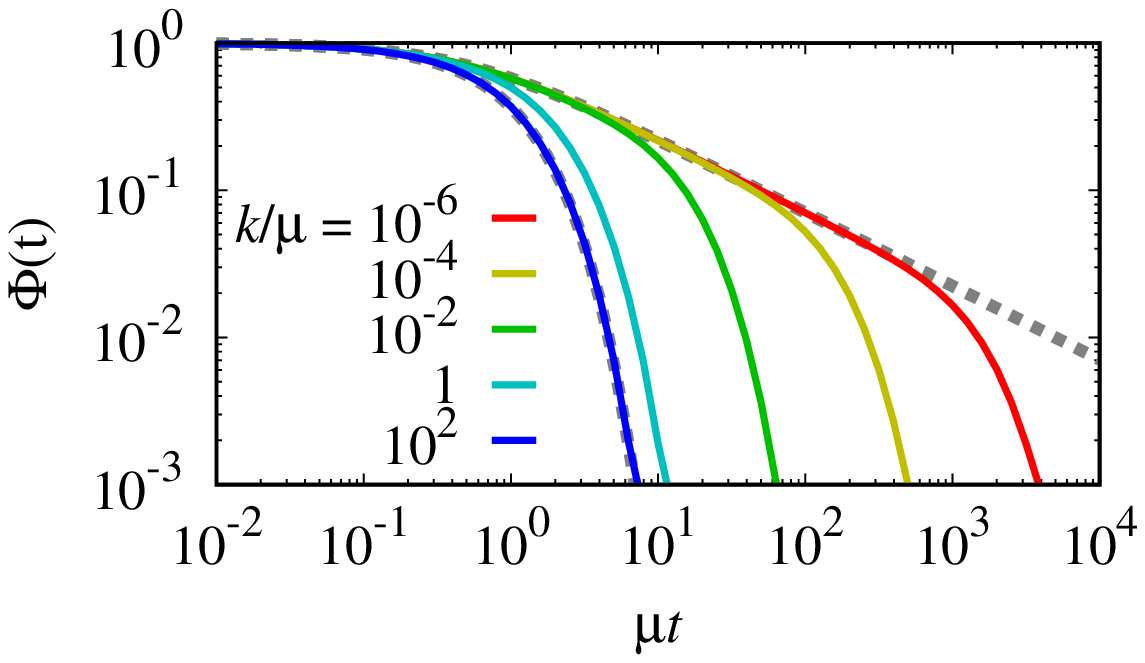}
\caption{\label{oufd_ou_relaxation_function}}
\end{figure}

We show the relaxation function for various values of $k / \mu$, directly
calculated by eq \eqref{relaxation_function_noise_explicit}, in
Figure~\ref{oufd_ou_relaxation_function}.  For comparison, the asymptotic forms
for $k / \mu \gg 1$ and $k / \mu \ll 1$
(eqs~\eqref{relaxation_function_noise_coefficient_fast_transition_limit} and
\eqref{relaxation_function_noise_coefficient_slow_transition_limit}) are also
shown Figure~\ref{oufd_ou_relaxation_function}.  We observe that for
sufficiently large $k / \mu$, the relaxation function $\Phi(t)$ is well
approximated by the asymptotic form. On the other hand, for small $k / \mu$ such
as $k / \mu = 10^{-6}$, we observe the deviation from the asymptotic form. This
is because the longest relaxation time is finite for finite $k / \mu$, as shown
in eqs \eqref{longest_relaxation_time_explicit} and
\eqref{longest_relaxation_time_explicit_asymptotic}. In the relatively short
time region ($t \lesssim 1 / \sqrt{k \mu}$), the asymptotic form works well.

It is straightforward to show the relation between the two relaxation functions
$\Phi(t)$ and $\Psi(t)$ becomes almost the same as the case of the Markovian
two-state model
(eqs~\eqref{relaxation_function_phi_psi_two_state_model_fast_transition_limit}
and \eqref{relaxation_function_phi_psi_two_state_model_slow_transition_limit}).
This implies that the OUFD
satisfies the relations $\Psi(t) \approx \Phi^{2}(t) \approx \Phi(2t)$ and
$\Psi(t) \approx \Phi(2 t)$  for
sufficiently fast and slow transition rates, respectively. These approximate
relations are expected to be independent of the details of the dynamics for the
diffusivity.

\subsection{Comparison with Other Models}
\label{comparison_with_other_models}

It would be informative to compare the results of the OUFD, with other models. Here we
consider two other models. One is the generalized Langevin equation {(GLE)}
model which is obtained by the projection operator method\cite{Evans-Morris-book} and widely
utilized to describe the dynamics of coarse-grained variables. Another
is the multi-mode OU process in which multiple
OU processes are linearly combined.
{
Currently, these conventional models are utilized as standard models to analyze
experimental data.
The GLE is widely utilized to analyze diffusion behavior of particles in
viscoelastic behavior\cite{Mason-Weitz-1995,Waigh-2005}.
The mutli-mode OU process is also widely utilized to analyze and model
non-single-exponential relaxation functions\cite{Larson-book}.
However, from the viewpoint of the FD, the analyses based on these
conventional models may not be physically reasonable for some systems.
Thus it would be informative to compare the properties of relaxation
functions in conventional models with those in the OUFD.}

First, we consider the GLE
with the memory function (the generalized OU process). In the generalized
OU process, the dynamic equation for $\bm{r}(t)$ is
described as a linear GLE.
For simplicity, we assume that the
dynamics is isotropic and the memory kernel can be expressed as a scalar quantity.
We can describe the dynamic equation as
\begin{equation}
 \label{linear_generalized_langevin_equation}
 \frac{d\bm{r}(t)}{dt}
  = - \alpha \bar{D} \int_{-\infty}^{t} dt' \, K(t - t') \bm{r}(t')
  + \sqrt{\bar{D}} \bm{\eta}(t) ,
\end{equation}
where $\bar{D}$ is constant (the reference diffusion coefficient), $K(t)$ is the
memory kernel, and $\bm{\eta}(t)$ is the Gaussian colored noise. The
fluctuation-dissipation relation of the second kind requires the noise to
satisfy the following relations:
{ $\langle \bm{\eta}(t) \rangle = 0$ and
$\langle \bm{\eta}(t) \bm{\eta}(t') \rangle = K(|t - t'|) \bm{1}$.}
The memory kernel can be simply related to the two point correlation
function (which is proportional to the relaxation function
$\Phi(t)$). Fox\cite{Fox-1977} showed that the memory kernel satisfies the following relation:
\begin{equation}
 \label{fox_relation}
 \tilde{\Phi}(u) = \frac{1}{u + \alpha \bar{D} \tilde{K}(u)} ,
\end{equation}
where $\tilde{\Phi}(u)$ and $\tilde{K}(u)$ are the Laplace transforms of the
relaxation function $\Phi(t)$ and the memory kernel $K(t)$:
{
 $\tilde{\Phi}(u) \equiv \int_{0}^{\infty} dt \, e^{-u t} \Phi(t)$ and
 $\tilde{K}(u) \equiv \int_{0}^{\infty} dt \, e^{-u t} K(t) $.
}
From eq~\eqref{fox_relation}, we can tune the memory kernel and reproduce the
relaxation function $\Phi(t)$ by the OUFD [eq~\eqref{relaxation_function_markovian_phi_eigenmodes}]. For
example, in the case of the Markovian two-state model, the Laplace transforms of
the relaxation function are calculated as follows, from
eq~\eqref{relaxation_function_phi_two_state_model}:
\begin{equation}
 \label{relaxation_function_phi_two_state_model_laplace_transform}
 \tilde{\Phi}(u) = \frac{\phi_{-}}{\lambda_{-} + u}
  + \frac{\phi_{+}}{\lambda_{+} + u} .
\end{equation}
From eqs~\eqref{fox_relation} and
\eqref{relaxation_function_phi_two_state_model_laplace_transform},
we find that the following Laplace-transformed memory kernel reproduces the the same relaxation
function as the Markovian two-state model:
\begin{equation}
 \tilde{K}(u) = \frac{1}{\alpha \bar{D}}
  \left[  (\phi_{-} \lambda_{-}  + \phi_{+} \lambda_{+}) -
 \frac{ \phi_{-} \phi_{+} (\lambda_{+} - \lambda_{-})^{2}
 } {u + (\phi_{-} \lambda_{+} + \phi_{+} \lambda_{-})} \right] .
\end{equation}
By performing the inverse Laplace transform, we have the following
simple expression in the time domain:
\begin{equation}
 \label{memory_kernel_generalized_langevin_equation}
 K(t) = \frac{1}{\alpha \bar{D}}
 \left[  (\phi_{-} \lambda_{-}  + \phi_{+} \lambda_{+}) \delta(t) -
 \phi_{-} \phi_{+} (\lambda_{+} - \lambda_{-})^{2} e^{- (\phi_{-}
   \lambda_{+} + \phi_{+} \lambda_{-}) t} \right] .
\end{equation}

However, thus obtained GLE
(eqs~\eqref{linear_generalized_langevin_equation} with
\eqref{memory_kernel_generalized_langevin_equation})
cannot reproduce
the relaxation function $\Psi(t)$ by the OUFD. 
The linear GLE \eqref{linear_generalized_langevin_equation} gives $\bm{r}(t)$ as a Gaussian
process. By utilizing the Wick's theorem, the multi point correlation functions of $\bm{r}(t)$
can be decomposed into two point correlation functions. For the case of
the four point correlation, which appears in the relaxation function
function $\Psi(t)$, we have
\begin{equation}
 \label{wick_decomposition_generalized_langevin}
\begin{split}
 & \langle [r_{x}(t) r_{y}(t)] [r_{x}(0) r_{y}(0)] \rangle \\
 & = \langle r_{x}(t) r_{y}(t) \rangle \langle r_{x}(0) r_{y}(0) \rangle
 + \langle r_{x}(t) r_{x}(0) \rangle \langle r_{y}(t) r_{y}(0) \rangle
 + \langle r_{x}(t) r_{y}(0) \rangle \langle r_{y}(t) r_{x}(0)
 \rangle \\
 & = \langle r_{x}(t) r_{x}(0) \rangle \langle r_{y}(t) r_{y}(0) \rangle .
\end{split}
\end{equation}
In the last line of eq~\eqref{wick_decomposition_generalized_langevin},
we have utilized the fact that the system is isotropic and there is no
correlation between $r_{x}(t)$ and $r_{y}(t)$.
From eq~\eqref{wick_decomposition_generalized_langevin},
the relaxation function $\Psi(t)$ is simply given as
\begin{equation}
 \label{relaxation_function_psi_generalized_langevin}
 \Psi(t) = \Phi^{2}(t) .
\end{equation}
Eq~\eqref{relaxation_function_psi_generalized_langevin} means that the
relaxation functions $\Phi(t)$ and $\Psi(t)$ contain essentially the same
information.  Here it should be stressed that
eq~\eqref{relaxation_function_psi_generalized_langevin} holds for any kernel
functions. In the case of Markovian $N$-state model,
eq~\eqref{relaxation_function_psi_generalized_langevin} generally consists of $N
(N + 1) / 2$ relaxation modes.  This is clearly different from the case of the
OUFD, where we have only $N$
relaxation modes for $\Psi(t)$.  Therefore we conclude that the
OUFD cannot be expressed as
the GLE. The effects of an FD seem
to be clearly observed when we analyze higher order correlation functions.  This
is consistent with the fact that the Langevin equation with an FD in absence of the potential exhibits only the normal diffusion
behavior on average, and the effects of an FD are observed in the
higher order fluctuations\cite{Uneyama-Miyaguchi-Akimoto-2015}.

Next, we consider the multi-mode OU process. We express
the position $\bm{r}(t)$ as the sum of $N$ modes. If we express the
$k$-th mode as $\bm{r}_{k}(t)$, $\bm{r}$ is expressed as the (weighted) sum of $\bm{r}_{k}$:
\begin{equation}
 \label{multi_mode_ornstein_uhlenbeck}
 \bm{r}(t) = \sum_{k = 1}^{N} c_{k} \bm{r}_{k}(t) ,
\end{equation}
where $c_{k}$ represents the weight factor for the $k$-th mode, and
$c_{k}$ is normalized to satisfy $\sum_{k = 1}^{N} c_{k}^{2} = 1$.
We assume that modes are statistically independent and
each mode obeys an OU process,
\begin{equation}
 \label{ornstein_uhlenbeck_k_mode}
 \frac{d\bm{r}_{k}(t)}{dt} = - \alpha D_{k} \bm{r}_{k}(t) + \sqrt{2 D_{k}}
  \bm{w}_{k}(t) ,
\end{equation}
where $D_{k}$ is the diffusion coefficient of the $k$-th mode ($D_{k}$ is assumed to be
constant) and
$\bm{w}_{k}(t)$ is the Gaussian white noise.  $\bm{w}_{k}(t)$ satisfies the following relations:
{ $\langle \bm{w}_{k}(t) \rangle = 0$ and
$\langle \bm{w}_{k}(t) \bm{w}_{l}(t') \rangle = \delta_{k l} \bm{1} \delta(t - t')$.}

Eq~\eqref{ornstein_uhlenbeck_k_mode} is just a usual
OU process and thus the relaxation function $\Phi(t)$
can be calculated straightforwardly. The two time correlation function
is calculated to be
{ $\langle \bm{r}_{k}(t) \cdot \bm{r}_{l}(0) \rangle = \delta_{kl} \langle \bm{r}_{k}^{2}(0) \rangle e^{-\alpha D_{k} t}$,}
and thus the relaxation function $\Phi(t)$ becomes
\begin{equation}
 \label{relaxation_function_phi_multi_mode_ornstein_uhlenbeck}
 \Phi(t) = \sum_{k = 1}^{N} c_{k}^{2} e^{- \alpha D_{k} t} .
\end{equation}
To reproduce the correlation function $\Phi(t)$ in the
Markovian two-state model, we simply set $N = 2$ and
{ then we have $\alpha D_{1} = \lambda_{-}$, $\alpha D_{2} = \lambda_{+}$,
$c_{1} = \phi_{-}^{1/2}$, and $c_{2} = \phi_{+}^{1/2}$.}

As the case of the GLE, even if we employ thus determined parameters, the multi-mode
OU process cannot reproduce the correlation function
$\Psi(t)$ correctly.
Due to the Gaussian nature, the relaxation function $\Psi(t)$ in the multi-mode OU
process can be calculated in a similar
way to the case of the GLE. The result is
\begin{equation}
 \label{relaxation_function_psi_multi_mode_ornstein_uhlenbeck}
 \Psi(t) = \Phi^{2}(t),
\end{equation}
and eq~\eqref{relaxation_function_psi_multi_mode_ornstein_uhlenbeck} is just the
same as eq~\eqref{relaxation_function_psi_generalized_langevin}.  Therefore, the
situation is the same as the case of the generalized OU model
with the memory kernel. (Actually, the multi-mode OU is a
Gaussian process and it can be also expressed as the linear GLE.)  The relaxation function $\Psi(t)$ has $N (N + 1) / 2$ relaxation
modes in the multi-mode OU model, whereas there is $N$ modes in
the OUFD.  We conclude that
the OUFD cannot be
expressed as the multi-mode OU process.

From the discussions above, we conclude that the OU
process with an FD belongs to a different class of
dynamic equations compared with widely utilized stochastic dynamic
equation models such as the GLE.
{ (This conclusion is rather trivial, since two independent stochastic
processes are multiplicatively coupled in the OUFD, whereas the couplings of stochastic
processes in the GLE and the multi-mode OU process are additive.)}
The importance of an FD
is especially observed via higher order correlations.
{In analogy to the non-Gaussianity parameter for
diffusion processes\cite{Rahman-1964}, we can
introduce a simple yet useful quantity which distinguishes the
OUFD, from the linear GLE and the multi-mode OU process:
\begin{equation}
 \label{non_gaussianity_parameter_for_relaxation}
 A(t) = \frac{\Psi(t)}{\Phi^{2}(t)} - 1.
\end{equation}
This quantity becomes zero if a relaxation process can be
described by the linear GLE or the multi-mode OU process. Conversely, if it is
non-zero, that process cannot be described by popular conventional
models, whereas the OUFD can successfully describe it.
Although it would not be easy to experimentally observe two relaxation functions
$\Phi(t)$ and $\Psi(t)$ for the same system, combinations of two
relaxation functions [such as
eq~\eqref{non_gaussianity_parameter_for_relaxation}] enables us to investigate heterogeneous dynamics of
the systems.}
We consider that a
FD will be especially useful to model and/or analyze
dynamics and relaxation behavior in heterogeneous environments such as
supercooled liquids\cite{Yamamoto-Onuki-1998,Yamamoto-Onuki-1998a,Sillescu-1999}.
It will be also informative to apply the concept of the FD to analyze the single molecule dynamics of
proteins\cite{Yang-Luo-Karnchanaphanurach-Louie-Rech-Cova-Xun-Xie-2003,Hu-Hong-Smith-Neusius-Cheng-Smith-2015}.

\section{Conclusions}
\label{colcusions}

In this work, we studied the relaxation behavior of the OUFD.  We modeled the stochastic process with
a linear restoring force and a thermal noise, both are coupled to a
time-dependent and fluctuating diffusivity.  We showed that the relaxation
functions $\Phi(t)$ and $\Psi(t)$ are expressed in terms of the integral of the
diffusion coefficient over time
[eqs~\eqref{relaxation_function_phi} and
  \eqref{relaxation_function_psi}]. To calculate the explicit forms of a
relaxation function, we utilized the functional integral expression with the
action functional and the transfer matrix method. We derived the simple
expression for the relaxation function, as the sum of relaxation modes
[eq~\eqref{relaxation_function_markovian_phi_eigenmodes}]. The
relaxation rate and the intensity of each mode is calculated from the eigenvalue
and eigenfunction of the transfer matrix.

As analytically solvable models, we studied the Markovian two-state model and
the OU type model for the noise coefficient. The two-state model
has only two relaxation modes, but the relaxation modes and relaxation
intensities depend both on the relaxation rates and the transition rates. This
is because the relaxation behavior of the OUFD is determined as a result of the competition between the
relaxation dynamics and the transition dynamics. If the transition rates are
sufficiently larger or smaller than the relaxation rates,
the corresponding relaxation function reduces to a simple asymptotic
form.
The situation is similar to the case when the dynamics for the noise coefficient is
described by the OU process.
In this model, the noise coefficient is a continuum
stochastic variable, and we have infinite relaxation modes. If the transition
rate is sufficiently smaller than the relaxation rate, we showed that
the relaxation function exhibits a power-law type behavior.
Because the relation between two relaxation functions $\Phi(t)$ and $\Psi(t)$ is not simple as in the cases
of the GLE and the multi-mode OU
process, we conclude that the OUFD is qualitatively different from those conventional models.
Thus, it is important and possible to unravel the underlying dynamics by
analyzing the two relaxation functions.

We believe that our model and analyses would be useful to
analyze some experimental data for supercooled liquids, polymers, and proteins. Now the authors
are working on the formulation of the relaxation function from the view
point of the renewal theory, and the
extension of this work will be published in the future\cite{Miyaguchi-inpreparation}.

\section*{Acknowledgment}
\label{acknowledgment}

T.U. was supported by Grant-in-Aid (KAKENHI) for Scientific Research C
JP16K05513.  T.M. was supported by Grant-in-Aid
  (KAKENHI) for Scientific Research C JP18K03417. 
T.A. was supported by Grant-in-Aid (KAKENHI)
for Scientific Research B JP16KT0021, and
Scientific Research C JP18K03468.

\appendix

{
\section{Relaxation Modes for Two-State Model}
\label{relaxation_modes_for_two_state_model}

In this appendix, we show the calculations for the two-state model in
Sec.~\ref{two_state_model}. From eq
\eqref{master_equation_diffusivity_two_state_model},
the transfer operator can be expressed as a $2 \times 2$ matrix form:
\begin{equation}
 \label{transfer_operator_two_state_model}
 \hat{\mathcal{W}} =
 \begin{bmatrix}
  \mu_{f} + k_{f} & - k_{s} \\
  - k_{f} & \mu_{s} + k_{s}
 \end{bmatrix} ,
\end{equation}
where we have expressed the relaxation rate at each state as
$\mu_{h} = \alpha D_{h}$ ($h = f, s$). The eigenvalues of the matrix in eq
\eqref{transfer_operator_two_state_model} is obtained as
\begin{equation}
 \label{eigenvalues_two_state_model_explicit}
 \lambda_{\pm} =
  \frac{1}{2}
  \left[ \mu_{f} + \mu_{s} + k_{f} + k_{s}
  \pm \sqrt{(\mu_{f} + \mu_{s} + k_{f} + k_{s})^{2}
  - 4 (k_{f} \mu_{s} + k_{s} \mu_{f} + \mu_{f} \mu_{s})}
  \right] .
\end{equation}
The first and second eigenvalues correspond to $\lambda_{-}$ and $\lambda_{+}$,
respectively.  We describe the second and first eigenvectors as
$\bm{\psi}_{\pm} = [\psi_{\pm,f}, \psi_{\pm,s}]^{\mathrm{T}}$. The eigenvectors
are calculated to be
\begin{equation}
 \label{eigenvectors_two_state_model}
  \bm{\psi}_{\pm} 
 =
 \begin{bmatrix}
  \lambda_{\pm} - \mu_{s} \\
  \mu_{f} - \lambda_{\pm}
 \end{bmatrix}  .
\end{equation}

From eq~\eqref{eigenvectors_two_state_model}
the relaxation function can be expressed as follows,
with thus obtained eigenvalues and eigenvectors
[eqs~\eqref{eigenvalues_two_state_model_explicit} and
\eqref{eigenvectors_two_state_model}] and the equilibrium distribution
[eq~\eqref{equilibrium_distribution_two_state_model}]:
\begin{equation}
 \label{relaxation_function_phi_two_state_model_explicit}
  \Phi(t)  =
  \begin{bmatrix}
   1 & 1
  \end{bmatrix}
  \cdot
  \begin{bmatrix}
   \bm{\psi}_{-} & \bm{\psi}_{+}
  \end{bmatrix}
  \cdot
  \begin{bmatrix}
   e^{-t \lambda_{-}} & 0 \\
   0 & e^{-t \lambda_{+}}
  \end{bmatrix}
  \cdot
  \begin{bmatrix}
   \bm{\psi}_{-} & \bm{\psi}_{+}
  \end{bmatrix}^{-1}
  \cdot
  \begin{bmatrix}
   P_{\text{eq},f} \\
   P_{\text{eq},s}
  \end{bmatrix} .
\end{equation}
Eq~\eqref{relaxation_function_phi_two_state_model_explicit} can be rewritten in a
simple form as $\Phi(t) = \phi_{-} e^{-\lambda_{-} t} + \phi_{+} e^{-
\lambda_{+} t}$, if we introduce the relaxation intensities defined as:
\begin{align}
 \label{intensity_two_state_model_minus_explicit}
 \phi_{-} & = \frac{1}{\lambda_{+} - \lambda_{-}} 
\left[ \lambda_{+} - \frac{  
    k_{f} \mu_{s} + k_{s} \mu_{f} }{k_{s} +
 k_{f}} \right], \\
 \label{intensity_two_state_model_plus_explicit}
 \phi_{+} & = \frac{1}{\lambda_{+} - \lambda_{-}} \left[  \frac{ 
    k_{f} \mu_{s} +  k_{s} \mu_{f} }{k_{s} +
 k_{f} } -
 \lambda_{-} \right] .
\end{align}
Thus we have
eq~\eqref{eigenvalues_two_state_model} and \eqref{intensity_two_state_model} in the main text.
}

{
\section{Relaxation Modes for Ornstein-Uhlenbeck Type Model}
\label{relaxation_modes_for_ornstein_uhlengeck_type_model}

In this appendix, we show the calculations for the OU type model in
Sec.~\ref{ornstein_uhlenbeck_type_model}. First we convert
eqs~\eqref{ornstein_uhlenbeck_noise_coefficient}
into the Fokker-Planck
equation.
Following a standard procedure\cite{vanKampen-book}, we have the following
Fokker-Planck equation for the distribution function of $b$, $P(b,t)$:
\begin{equation}
 \label{fokker_planck_equation_noise_coefficient}
 \frac{\partial P(b,t)}{\partial t} = \hat{\mathcal{L}} P(b,t) ,
\end{equation}
with the Fokker-Planck operator defined as
\begin{equation}
 \label{fokker_planck_operator_noise_coefficient}
 \hat{\mathcal{L}} P = k \frac{\partial}{\partial b}
  \left[ b P + \frac{\partial P}{\partial b} \right] .
\end{equation}
Obviously, the equilibrium distribution $P_{\text{eq}}(b)$ is a
Gaussian:
$P_{\text{eq}}(b) = e^{- b^{2} / 2}  / \sqrt{2 \pi}$.

From eqs~\eqref{transfer_operator_explicit} and
\eqref{fokker_planck_operator_noise_coefficient}, the transfer operator can be
explicitly expressed as
\begin{equation}
 \label{transfer_operator_noise_coefficient}
 - \hat{\mathcal{W}} \psi
  = - \mu b^{2} \psi + k \frac{\partial }{\partial b}
  \left[ b \psi + \frac{\partial \psi}{\partial b} \right] ,
\end{equation}
where $\mu \equiv \alpha \bar{D}$.  The eigenvalue $\lambda$ and the
eigenfunction $\psi(b)$ satisfy
the eigenvalue equation,
\begin{equation}
 \hat{\mathcal{W}} \psi(b) = \lambda \psi(b) .
\end{equation}
Here we introduce the variable transform to make the transfer operator
self-adjoint\cite{Risken-book}:
\begin{equation}
 \label{variable_transform_engenfunction_noise_coefficient}
 \tilde{\psi}(b) \equiv e^{b^{2} / 4} \psi(b)
 = e^{- b^{2} / 4} \psi^{\dagger}(b).
\end{equation}
Then we have the following eigenvalue equation:
\begin{equation}
 \label{eigenvalue_equation_noise_coefficient}
  \gamma^{2} b^{2} \tilde{\psi}
   - \frac{d^{2} \tilde{\psi}}{d b^{2}}
   = \left(\frac{\lambda}{k} + \frac{1}{2} \right) \tilde{\psi},
\end{equation}
where $\gamma = \sqrt{\mu / k + 1 / 4}$.
Roughly speaking, the parameter $\gamma$ represents the competition between the
transition and relaxation. If the transition becomes faster than the relaxation,
$\gamma$ decreases. If the relaxation becomes faster than the transition,
$\gamma$ increases. It should be noticed that $\gamma$ satisfies $\gamma > 1 /
2$.  This eigenvalue equation has the same form as the Schr\"{o}dinger equation
for a one dimensional harmonic potential\cite{Schiff-book}, and thus we can
calculate eigenfunctions and eigenvalues straightforwardly.  The $n$-th
eigenvalue and eigenfunction ($n = 0,1,2,\dots$) are given as:
\begin{equation}
 \label{eigenvalue_noise_coefficient}
 \lambda_{n}
 = k [ (2 n + 1) \gamma - 1 /2 ] ,
\end{equation}
\begin{equation}
 \label{eigenfunction_noise_coefficient}
 \tilde{\psi}_{n}(b) =
 \left(\frac{\gamma^{1/2}}{\pi^{1/2} 2^{n} n!}\right)^{1/2} H_{n}(\sqrt{\gamma} b) e^{-\gamma b^{2} /
  2} ,
\end{equation}
where $H_{n}(x)$ is the $n$-th order Hermite polynomial\cite{NIST-handbook,Schiff-book}.

The relaxation function can be expressed with the eigenvalues and
eigenfunctions. From eqs~\eqref{variable_transform_engenfunction_noise_coefficient},
\eqref{eigenvalue_noise_coefficient}, and
\eqref{eigenfunction_noise_coefficient},
we have
\begin{equation}
 \label{relaxation_function_noise_coefficient_explicit}
 \Phi(t) = \sum_{n = 0}^{\infty} \phi_{n}
 \exp\left[ - k  [ (2 n + 1) \gamma - 1 /2 ] t \right] ,
\end{equation}
with
\begin{equation}
 \label{intensity_noise_coefficient}
  \phi_{n}
  = \frac{\gamma^{1/2}}{\pi 2^{n + 1/2} n!}
  \left[ \int_{-\infty}^{\infty} db \,
  H_{n}(\sqrt{\gamma} b) e^{-\gamma b^{2} /  2 - b^{2} / 4} \right]^{2} .
\end{equation}

The integral in eq~\eqref{intensity_noise_coefficient} can be calculated analytically.
}
%
%
We rewrite eq~\eqref{intensity_noise_coefficient} as
\begin{equation}
 \label{intensity_noise_coefficient_modified}
  \phi_{n}
  = \frac{\gamma^{-1/2}}{\pi 2^{n + 1/2} n!} I_{n}^{2} ,
\end{equation}
where $I_{n}$ is the following integral:
\begin{equation}
 \label{integral_i_n}
  I_{n} \equiv \int_{-\infty}^{\infty} ds \,
 H_{n}(s) e^{-s^{2} /  2 -  s^{2} / 4 \gamma} .
\end{equation}
For $n = 0$, the integral $I_{0}$ can be easily calculated, because $H_{0}(s) =
1$. We have
\begin{equation}
 \label{integral_i_0}
  I_{0} = \int_{-\infty}^{\infty} ds \, e^{- s^{2} /  2 - s^{2} / 4 \gamma} 
  = \sqrt{\frac{2 \pi \gamma}{\gamma + 1 / 2}} .
\end{equation}
Also, the integral $I_{n}$ can be calculated easily for odd $n$. In this case,
from the symmetry of the Hermite polynomial, $H_{n}(s) = - H_{n}(s)$,
the integrand in eq~\eqref{integral_i_n} is an odd function of $s$. Then, we simply have
\begin{equation}
 \label{integral_i_n_odd}
  I_{n} = 0 \qquad (n : \text{odd}), 
\end{equation}
and thus the odd $n$ modes vanish:
\begin{equation}
 \label{intensity_noise_coefficient_odd}
  \phi_{n} = 0 \qquad (n : \text{odd}).
\end{equation}

Thus now we need to calculate the integral $I_{n}$ for even $n \ge 2$.
By utilizing the recurrence relation for the Hermite polynomial\cite{NIST-handbook,Schiff-book},
\begin{equation}
 H_{n + 1}(s) = 2 s H_{n}(s) - 2 n H_{n - 1}(s),
\end{equation}
we have
\begin{equation}
 \begin{split}
  I_{n} & =  \int_{-\infty}^{\infty} ds \,
  [2 s H_{n - 1}(s) - 2 (n - 1) H_{n - 2}(s)] e^{-s^{2} /  2 -  s^{2} / 4 \gamma} \\
  & = 2 \int_{-\infty}^{\infty} ds \,
  s H_{n - 1}(s)  e^{-s^{2} /  2 -  s^{2} /  4 \gamma}
  - 2 (n - 1) I_{n - 2} \\
  & = \frac{2}{1 + 1 / 2 \gamma} \int_{-\infty}^{\infty} ds \,
  \frac{dH_{n - 1}(s)}{ds}  e^{-s^{2} /  2 -  s^{2} /  4 \gamma}
  - 2 (n - 1) I_{n - 2} .
 \end{split}
\end{equation}
In the last line, we have utilized the partial integral. We utilize
another recurrence relation for the Hermite polynomial\cite{NIST-handbook,Schiff-book}:
\begin{equation}
 \frac{dH_{n}(s)}{ds} = 2 n H_{n - 1}(s) .
\end{equation}
Finally we have the following recursive relation for the integral $I_{n}$:
\begin{equation}
 \label{integral_i_n_recursive_relation}
 \begin{split}
  I_{n}
  & = \frac{4 (n - 1)}{1 + 1 / 2 \gamma} I_{n - 2}
  - 2 (n - 1) I_{n - 2} \\
  & = 2 (n - 1) \frac{\gamma - 1 / 2}{\gamma + 1 / 2} I_{n - 2} ,
 \end{split}
\end{equation}
and from eqs~\eqref{integral_i_0} and
\eqref{integral_i_n_recursive_relation}, the solution is
\begin{equation}
 \label{integral_i_n_final}
 \begin{split}
  I_{n} & = 2^{n / 2} (n - 1)!!
  \left(\frac{\gamma - 1/2}{\gamma +  1/2}\right)^{n/2} I_{0} \\
  & = \sqrt{2 \pi} 2^{n / 2} (n - 1)!!
  \frac{\gamma^{1/2} (\gamma - 1/2)^{n/2}}{(\gamma +  1/2)^{n/2 + 1/2}} .
 \end{split}
\end{equation}
Here, $n!!$ represents the double factorial of $n$\cite{Arfken-Weber-Harris-book}.

By substituting eq~\eqref{integral_i_n_final} into
\eqref{intensity_noise_coefficient_modified}, and utilizing some
relations for the double factorial and the factorial\cite{Arfken-Weber-Harris-book}, the intensity of the
$n$-th mode is explicitly expressed as
\begin{equation}
 \label{intensity_noise_coefficient_final}
   \phi_{n}
    = \frac{\gamma^{-1/2}}{\pi 2^{n + 1/2} n!} I_{n}^{2}  
    = \frac{\sqrt{2} n!}{2^{n} [(n / 2)!]^{2}} 
  \frac{\gamma^{1/2}(\gamma - 1/2)^{n}}{(\gamma +  1/2)^{n + 1}} .
\end{equation}

{ Eqs~\eqref{relaxation_function_noise_coefficient_explicit},
\eqref{intensity_noise_coefficient_odd} and
\eqref{intensity_noise_coefficient_final} give eqs~\eqref{relaxation_function_noise_coefficient}
and \eqref{intensity_noise_coefficient_explicit} in the main text.}

{
\section{Slow Transition Limit of Ornstein-Uhlenbeck Type Model}
\label{calculation_of_relaxation_function_at_slow_transitoin_limit}

In this appendix, we show the detailed calculation of the relaxation
function of the OU type model, at the slow transition limit where $k \ll \mu$.
As we mentioned in the main text, it is difficult to obtain an approximate form eq~\eqref{relaxation_function_noise_coefficient}.
Instead, here we approximate the transfer operator and solve the eigenvalue
equation with the approximate transfer operator. For $k
\ll \mu$, the transfer
operator [eq~\eqref{transfer_operator_noise_coefficient} in Appendix~\ref{relaxation_modes_for_ornstein_uhlengeck_type_model}]
can be approximated as
\begin{equation}
 \label{transfer_operator_noise_coefficient_slow_transition_limit}
 - \hat{\mathcal{W}} \psi \approx - \mu b^{2} \psi .
\end{equation}
From eq~\eqref{transfer_operator_noise_coefficient_slow_transition_limit},
the eigenvalue equation
[eq~\eqref{eigenvalue_equation_noise_coefficient} in
Appendix~\ref{relaxation_modes_for_ornstein_uhlengeck_type_model}] can
be simply approximated as
\begin{equation}
 \label{eigenvalue_equation_noise_coefficient_slow_transition_limit}
 \lambda \psi(b) \approx \mu b^{2} \psi(b) .
\end{equation}
Eq \eqref{eigenvalue_equation_noise_coefficient_slow_transition_limit} is not a differential
equation unlike
eq~\eqref{eigenvalue_equation_noise_coefficient}. Formally, we have the
following eigenvalue and the eigenfunction for eq \eqref{eigenvalue_equation_noise_coefficient_slow_transition_limit}:
\begin{equation}
 \label{eigenvalue_noise_coefficient_slow_transition_limit}
 \psi(b,a) = \psi^{\dagger}(b,a) = \delta(b - a), \qquad \lambda(a) = \mu a^{2}
\end{equation}
where $a$ is the index of the eigenvalue and eigenfunction, and $a$ is a
continuum variable. (The eigenvalue is not in the ascending order in
$a$, but here we do not need the eigenvalues to be ordered thus we
simply use
eq~\eqref{eigenvalue_noise_coefficient_slow_transition_limit}.)
The
sum over all the eigenmodes should be replaced by the integral over the
index $a$.
Therefore, we have the following approximate expression for the
relaxation function $\Phi(t)$:
\begin{equation}
 \label{relaxation_function_noise_coefficient_slow_transition_limit_explicit}
\begin{split}
 \Phi(t) & \approx \int_{-\infty}^{\infty} da \, 
 \left[ \int_{-\infty}^{\infty} db \, \delta(b - a) \int_{-\infty}^{\infty} db' \, \delta(b' - a) \frac{1}{\sqrt{2 \pi}}
  e^{-b'^{2} / 2}  \right] e^{- \mu a^{2} t} \\
 & \approx \frac{1}{\sqrt{1 + 2 \mu t}} .
\end{split}
\end{equation}
Thus we have eq~\eqref{intensity_noise_coefficient_explicit}
in the main text.
}

\clearpage

\bibliographystyle{apsrev4-1}
\bibliography{relaxation_function_oufd}

\clearpage


\section* {Figure Captions}

\hspace{-\parindent}%
{ Figure~\ref{oufd_image}: An example of a realization of
the stochastic process which obeys the OUFD in one dimension, $x(t)$. The solid black curve represent
the position $x(t)$ at time $t$, and the background colors represents the
diffusivity $D(t)$. The position is fluctuating around the origin ($x = 0$, the
dashed black line) due to the restoring force.}

\

\hspace{-\parindent}%
Figure~\ref{oufd_two_state_relaxation_function}: The relaxation function
$\Phi(t)$ for the OUFD by
the Markovian two-state model.  (a) The relaxation rate is $\mu_{s} / \mu_{f} =
10^{-2}$ and the transition rate is changed. { The solid
curves represent data for
$\kappa = k_{f} / \mu_{f} = k_{s} / \mu_{f} = 10^{-2}, 10^{-1}, 1,$ and
$10^{1}$}. The dotted gray curves represent asymptotic
forms [eqs~\eqref{relaxation_function_phi_two_state_model_fast_transition_limit}
and \eqref{relaxation_function_phi_two_state_model_slow_transition_limit}].  (b)
The transition rate is constant $\kappa = 10^{-2}$ and the relaxation rate is
changed.  { The solid curves represent data for $\mu_{s} / \mu_{f} = 1, 10^{-1},
10^{-2}$, and $10^{-3}$.}  The dotted gray curves represent
the asymptotic forms.

\

\hspace{-\parindent}%
Figure~\ref{oufd_ou_relaxation_function}:
The relaxation function $\Phi(t)$ for the OUFD. The noise coefficient obeys the
OU model. { The solid curves represent the
relaxation functions for $k / \mu = 10^{-6}, 10^{-4}, 10^{-2}, 1,$ and
$10^{2}$}, calculated by eq~\eqref{relaxation_function_noise_explicit}. The dotted gray curves represent the
asymptotic forms for $k / \mu \gg 1$ [left, 
eq~\eqref{relaxation_function_noise_coefficient_fast_transition_limit}] 
and $k / \mu \ll 1$ [right,
eq~\eqref{relaxation_function_noise_coefficient_slow_transition_limit}].

\clearpage

\section* {Figures}

\clearpage

\clearpage

\end{document}